\documentclass[twocolumn,prd,aps,tightenlines,floats,floatfix,eqsecnum,preprintnumbers,nofootinbib]{revtex4}

\def\sp{\kern +3pt}
\def\sm{\kern -3pt}
\def\spQ{\kern +6pt}

\def\bea{\begin{eqnarray}}
\def\eea{\end{eqnarray}}

\def\sfrac#1#2{{\textstyle \frac{#1}{#2}}}

\def\be{\begin{equation}}
\def\ee{\end{equation}}
\def\ba{\begin{eqnarray}}
\def\ea{\end{eqnarray}}

\usepackage{graphics}
\usepackage{graphicx}
\usepackage{epsf}
\usepackage{amsmath}
\usepackage{amssymb}

\usepackage{color}

\begin{document}

\phantom{0}
\vspace{-0.2in}
\hspace{5.5in}

\preprint{  }

\vspace{-1in}

\title
{\bf 
Semirelativistic approximation 
to the  $\gamma^\ast N \to N(1520)$ \\
and  $\gamma^\ast N \to N(1535)$ transition
form factors}
\author{G.~Ramalho}
\vspace{-0.1in}

\affiliation{
International Institute of Physics, Federal
University of Rio Grande do Norte, 
Campus Universit\'ario -- Lagoa Nova, CP.~1613, 
Natal, Rio Grande do Norte 59078-970, Brazil
}

\vspace{0.2in}
\date{\today}

\phantom{0}

\begin{abstract}
The representation of the wave functions 
of the nucleon resonances within a relativistic framework 
is a complex task.
In a  nonrelativistic framework 
the orthogonality between states can be imposed naturally.
In a relativistic generalization, however, 
the derivation of the orthogonality condition between states
can be problematic, particularly  when the states have different masses.  
In this work we study the  $N(1520)$ and $N(1535)$ states 
using a relativistic framework.
We considered wave functions derived in previous works,
but impose the orthogonality between the nucleon 
and resonance states using the properties of the nucleon,
ignoring the difference of masses between the states
(semirelativistic approximation).
The $N(1520)$ and $N(1535)$ wave functions are then 
defined without any adjustable parameters
and are used to make predictions for the valence quark contributions 
to the transition form factors.
The predictions compare well with the data 
particularly for high momentum transfer, 
where the dominance of the quark degrees of freedom is expected.
\end{abstract}

\vspace*{0.9in}  
\maketitle

\section{Introduction}

In the last century we learned that 
the hadrons, including  the nucleon ($N$)
and the nucleon 
excitations ($N^\ast$) are not pointlike particles 
and have their own internal structure.
The structure of those states 
is the result of the internal constituents, quarks and gluons, 
and the interactions ruled by Quantum Chromodynamics (QCD).  
In the last decades experimental facilities
such as Jefferson Lab (JLab), MAMI (Mainz) 
and MIT-Bates have accumulated information (data) 
about the electromagnetic structure of 
the nucleon resonances, parametrized in terms of structure form factors 
for masses up to 3 GeV~\cite{Aznauryan12a,NSTAR}.

Several theoretical models
have been proposed to interpret the nucleon resonance spectrum  
and the information associated with 
its internal structure~\cite{Aznauryan12a,NSTAR,Capstick00}.
Different models provide different 
parametrizations of the internal structure 
in terms of the effective degrees of freedom.
Some of the more successful models 
are the constituent quark models (CQM) 
based on nonrelativistic kinematics like the 
Karl-Isgur model~\cite{Karl-Isgur,Capstick00} 
and the Light Front quark models 
(LFQM) defined 
in the infinite momentum frame~\cite{Capstick95,Aznauryan07,Aznauryan12b}.
In those extreme cases, nonrelativistic models or LFQM,
the kinematics is simplified.
In general, however, the transition 
between the  nonrelativistic and relativistic 
regimes is not a trivial task.

In this work we discuss the $\gamma^\ast N \to N^\ast$
transition form factors for resonances $N^\ast$  
with negative parity.
The definition of the wave functions of 
the nucleon (mass $M_N$) and a nucleon excitation (mass $M_R$),
in terms of the internal quark degrees of freedom, 
can be done first in the rest frame of the particle, 
and extended later for a moving frame
using a Lorentz transformation.
In a nonrelativistic framework the mass and energy 
of the state are not relevant for the definition 
of the states.
Moreover, the orthogonality between the nucleon and the resonance
$N^\ast$ is ensured, 
since the wave functions are independent of their masses.
To understand the complexity of the generalization of 
the orthogonality condition 
from a nonrelativistic framework to a relativistic framework, 
we consider the example of charge operator, $J^0$,
for a transition between the  nucleon ($N$) and a 
 spin 1/2 negative parity state ($R$).
The projection of $J^0$ in the nucleon and $N^\ast$ states at  
zero square momentum transfer ($q^2=0$)
is proportional to the overlap between wave functions. 
One can show that in a relativistic framework 
the overlap is proportional 
to $\bar u_R \gamma_5 u_N$~\cite{S11}.
In a framework where we can neglect the mass difference 
between the states,\footnote{
The nonrelativistic limit 
can also be defined as the equal mass limit ($M_R=M_N$)
or as the heavy baryon limit, when 
the terms on   
$(M_R -M_N)/M_N$  can be neglected.}
we obtain $\bar u_R \gamma_5 u_N = 0$,
for the case where $N$ and $R$ have the same momentum ($q^2=0$).
We then conclude that in the nonrelativistic limit
the orthogonality between states is naturally ensured.
In a relativistic framework 
the imposition of the orthogonality condition is more complex, 
since the nucleon and the resonance $R$ cannot be at rest in the same frame,
and the boost changes the properties of the states.
As a consequence, states that are orthogonal 
when the mass difference can be neglected may not be orthogonal 
when the mass difference is taken into account.

The problem of how to define a wave function of 
a nucleon excitation that generalizes 
the nonrelativistic structure of the state 
and is also orthogonal to the nucleon  
was already discussed in the context 
of the covariant spectator quark model 
for the negative parity resonances $N(1520)$
and $N(1535)$~\cite{D13,S11,SQTM}.
The solution at the time was to 
define the radial wave functions for the $N^\ast$ states 
in order to ensure the orthogonality with the nucleon state.
The price to pay was the introduction of 
a new momentum scale parameter in the radial wave functions, 
to be determined by the phenomenology.

In this work we discuss an alternative approach.
Instead of focusing on the necessity of imposing the orthogonality between states, 
we assume that the mass difference is not the 
more relevant factor and treat 
the two states as different states with the same mass, $M$,
defined by the average $M= \frac{1}{2}(M_N + M_R)$.
We then consider wave functions 
of states with the same mass.
We call this approximation 
the semirelativistic approximation, since 
it keeps the features of the nonrelativistic regime 
(no mass dependence) and preserves 
the covariance of the states.

The great advantage of the previous assumption 
is that, as explained in detail later, 
one can relate the radial wave function
of the resonance $R$ with the radial wave function 
of the nucleon, increasing the predictive power of the model.
We use the semirelativistic approximation to
calculate transition form factors for the resonances 
$N(1520)$ and $N(1535)$.
The calculation of the helicity amplitudes
is more problematic since their relation 
with the form factors depends on 
the nucleon and resonance physical masses.
Later on, we discuss how to calculate the helicity amplitudes
using the form factors defined in the equal mass limit.

In this work we show that, 
the results from the semirelativistic approximation
compare well with $\gamma^\ast N \to N(1520)$ 
and $\gamma^\ast N \to N(1535)$ form factor 
data~\cite{PDG,CLAS1,CLAS2,Mokeev16,MAID1,MAID2,Dalton09}, 
particularly for large square momentum transfer ($Q^2=-q^2$).
At low $Q^2$, the agreement is not so good, 
since the meson cloud contributions 
are expected to be important and even dominant in some 
transitions~\cite{Aznauryan12a,NSTAR,EBAC,EBACreview,Tiator04,Burkert04}.

This article is organized as follows:
In Sec.~\ref{secOrthogonality}, 
we discuss the orthogonality between states
and explain how the orthogonality 
can be imposed in a relativistic framework.
In Sec.~\ref{secFormalism}, we present 
the formalism associated with the $\gamma^\ast N \to N(1520)$
and  $\gamma^\ast N \to N(1535)$ transitions,
and the relations between electromagnetic currents, 
helicity amplitudes and electromagnetic form factors.
Next, in Sec.~\ref{secCSQM}, 
we discuss the covariant spectator quark model 
and present the model predictions for the transitions under study.
The results of the semirelativistic approximation 
are discussed in Sec.~\ref{secResults}.
Outlook and conclusions are presented in Sec.~\ref{secConclusions}.

\section{Orthogonality and relativity}
\label{secOrthogonality}

We discuss now the orthogonality between the nucleon 
and a nucleon excitation $R$.
Since they represents different systems 
they should be represented by orthogonal wave functions,
$\Psi_N$ and $\Psi_R$, respectively.
In a quark-diquark model one can express those 
wave functions as $\Psi_N(P,k)$ and $\Psi_R(P,k)$,
where $P$ and $k$ are respectively the baryon and the diquark momenta
($P-k$ is the momentum of the single quark).
For simplicity we ignore the indices associated 
with the angular momentum, the parity, and  
the spin and isospin projections.

In a nonrelativistic framework, the orthogonality 
between the wave functions is ensured when 
the overlap between the two wave functions vanishes 
in the limit where both particles have zero 
three-momentum, ${\bf P}={\bf 0}$, which can be represented, 
ignoring the isospin effect for now, by the condition
\ba
\sum_\Gamma \int_k \Psi_R^\dagger (\bar P,k) \Psi_N(\bar P,k)= 0,
\label{eqOrthNR}
\ea
In the previous equation $\Gamma$ is a 
diquark polarization index 
and $\bar P= (M,{\bf 0})$ is the nucleon and $R$ momenta
($\bar P$ is used to label $P$ in the limit $Q^2=0$).
The integral symbol represent the 
covariant integration over the diquark momentum.
The mass/energy component was introduced to facilitate 
the relativistic generalization, but it is irrelevant 
for the present discussion,
since in the nonrelativistic limit the wave functions 
are defined only in terms of the three-momentum.
It is important to note that in 
Eq.~(\ref{eqOrthNR}) the functions are defined 
for the zero three-momentum transfer 
($|{\bf q}|=0$),
since both states wave the same  three-momentum  ${\bf P}={\bf 0}$.
In a covariant language we can write $ Q^2= -|{\bf q}|^2 =0  $,
since we assumed that the energy is irrelevant 
for transitions at $|{\bf q}|^2=0$.

The question now is how to generalize the condition (\ref{eqOrthNR}),  
defined for $Q^2=0$, to the relativistic case, 
particularly for the unequal mass case.
In the context of the
covariant spectator quark model \cite{Nucleon,Nucleon2,Omega},
 the problem was already 
discussed for several baryon systems \cite{S11,D13,N1710,Roper,NDeltaD}.
In that formalism the relativistic generalization 
of Eq.~(\ref{eqOrthNR}) is
\ba
\sum_\Gamma \int_k \Psi_R^\dagger (\bar P_+,k) \Psi_N(\bar P_-,k)= 0,
\label{eqOrthRel}
\ea
where $\bar P_+$ and $\bar P_-$ represent 
the resonance $R$ and the nucleon momenta, respectively,
in the case $Q^2=0$.
Taking for instance the $R$ rest frame,
one has, 
assuming that the momentum transfer ${\bf q}$ is along 
the $z$-axis:
\ba
& &
\bar P_+= (M_R,0,0,0), \nonumber \\
& &
\bar P_- = \left(  E_N, 0,0,- |{\bf q}| \right),
\label{eqP0}
\ea
where $E_N= \frac{M_R^2+ M_N^2}{2M_R}$
and  $|{\bf q}|= \frac{M_R^2-M_N^2}{2M_R}$.

From the previous relations we conclude 
that in the case $Q^2=0$, we cannot 
have $R$ and $N$ at rest at the same time 
(in the same frame) unless $M_R=M_N$.
Thus,  in the conditions of Eqs.~(\ref{eqP0}), 
the resonance $R$ is at rest, 
but the nucleon is not at rest ($|{\bf q}| \ne 0$).

The discussion about the orthogonality between  
states that are not defined in the same rest frame
is more complex, and has consequences 
in the calculation of the transition form factors in the limit $Q^2 =0$.
We can illustrate the problem 
looking for the magnetic form factor $G_M$ for the 
$\gamma^\ast  N \to N(1520)$ transition.
As discussed in Refs.~\cite{D13},
the orthogonality condition implies 
that $G_M(0) \propto {\cal I}_R(0)$,
where ${\cal I}_R (Q^2)$ is a integral 
defined by the overlap between the nucleon and $R$ 
radial wave functions (the details can be found in Refs.~\cite{D13}).
Since the orthogonality condition between states 
is equivalent to ${\cal I}_R(0)=0$~\cite{D13}, 
one obtains $G_M(0)=0$,
in contradiction with the 
experimental result $G_M(0)= -0.393 \pm    0.044 $~\cite{PDG}.

In the framework of the covariant spectator quark model, 
we can prove that the orthogonality condition 
(\ref{eqOrthRel}) 
for the states $R=N(1520)$, $N(1535)$
is equivalent to~\cite{S11,D13} 
\ba
{\cal I}_R (0)  \equiv
\int_k \frac{k_z}{|{\bf k}|} \psi_R(\bar P_+,k) 
\psi_N(\bar P_-,k)=0,
\label{eqOrthRel2}
\ea
where $\psi_R$ and $\psi_N$ are radial wave functions
from $R$ and $N$, respectively
and real functions of $(\bar P_\pm-k)^2$.
The integral ${\cal I}_R(0)$ is defined 
in  Eq.~(\ref{eqOrthRel2}) at the $R$ rest frame, 
by simplicity.
The general expression can be found in Refs.~\cite{D13,S11}.

In Sec.~\ref{secCSQM}, we present 
the results for the  $\gamma^\ast N \to N(1520)$
and $\gamma^\ast N \to N(1535)$ form factors  
and the connection with the helicity amplitudes
within the covariant spectator quark model.

For the  $\gamma^\ast N \to N(1520)$ transition,
one has 3 independent form factors $G_i$  ($i=1,2,3$),
with the form~\cite{D13}
\ba
G_i (Q^2) \propto \frac{{\cal I}_R(Q^2)}{|{\bf q}|},
\label{eqGi}
\ea
when $Q^2 \to 0$.
In the case $\frac{{\cal I}_R(Q^2)}{|{\bf q}|} \to \mbox{const}$,
one has finite contributions for the transverse amplitudes, 
$A_{1/2}(0)$ and  $A_{3/2}(0)$, consistently with the data.

As for the  $\gamma^\ast N \to N(1535)$ transition, 
we conclude that the two 
independent form factors $F_i^\ast$ ($i=1,2$)
can be represented as \cite{S11}
\ba
F_i^\ast (Q^2) \propto {\cal I}_R(Q^2),
\label{eqFi}
\ea
when $Q^2 \to 0$.
In addition, it can be shown that ${\cal I}_R \propto |{\bf q}|$
when the nucleon and the $N(1535)$ states
are described by the same radial wave function \cite{S11}.
As a consequence of Eq.~(\ref{eqFi}), 
one obtains $F_1^\ast(0)=0$, 
automatically in the limit $|{\bf q}| \to 0$.

The problem associated with the results from the 
$\gamma^\ast N \to N(1520)$ and $\gamma^\ast N \to N(1535)$
form factors given by Eqs.~(\ref{eqGi})
and (\ref{eqFi}) is that they are finite only in the case $|{\bf q}| \to 0$
when $Q^2=0$, which is inconsistent with 
$|{\bf q}|= \frac{M_R^2-M_N^2}{2 M_R} \ne 0$, 
unless $M_R = M_N$.

In previous works~\cite{S11,D13}, 
we developed models that violate 
the orthogonality condition ~(\ref{eqOrthRel2}) 
as for the $\gamma^\ast N \to N(1535)$ transition~\cite{S11}, 
or are consistent with the orthogonality condition, but 
failed to describe the low $Q^2$ data, 
as for the $\gamma^\ast N \to N(1535)$ transition~\cite{D13}.

In the present work, we consider an alternative approach
that tries to achieve two goals.
On the one hand we want to keep the nice 
analytic properties of the 
form factors in the case $M_R=M_N$,
which are spoiled in the 
relativistic generalization of the wave function 
in the case $M_R \ne M_N$.
On the other hand, we want  
to describe the experimental 
helicity amplitudes, which are defined only 
in the case $M_R \ne M_N$. 
With those two ideas in mind we consider the 
following approximation: 
we assume that both states, 
the nucleon and the resonance $R$
are states with the same mass, given 
by the average between the two physical masses
\ba
M= \frac{1}{2}(M_R + M_N).
\ea 
With this choice, the  
orthogonality condition (\ref{eqOrthRel2}) 
is automatically ensured if $\psi_R$ is defined as $\psi_N$. 
In that case the product $\psi_R(\bar P_+,k) \psi_N(\bar P_-,k)$
is symmetric in the angular variables when $|{\bf q}|=0$, 
as a consequence the integral in $k_z$ vanishes.

Since this approximation mimics 
the nonrelativistic regime
when the mass difference is ignored,
we refer to this approximation as the semirelativistic approximation.

A nice consequence of the semirelativistic approach is that, 
since the $R$ states are defined using 
the radial wave function of the nucleon ($\psi_N$),
there are no adjustable parameters in the model.
Therefore, the results of the present model are true predictions 
that can be compared with the experimental data.

\section{Formalism}
\label{secFormalism}

In this section, we present the 
general definitions of  the 
$\gamma^\ast N \to R$ helicity amplitudes
at the final state ($R$) rest frame.
Following the notation of previous works 
we use $P_-$ for the initial state (nucleon)
and $P_+$ for the final state ($R$).
The momentum transfer is then \mbox{$q= P_+ -P_-$.}
We use also $Q^2=-q^2$, which  
we relabel as the square momentum transfer.
The transition current operator is represented by $J^\mu$,
and is defined in units of the proton charge $e$.
The explicit form of $J^\mu$ depends of the $N$ and $R$ states.
To express the projection of $J^\mu$ 
in the states $R$ and $N$ 
we use the matrix element
\ba
J^\mu_{NR} &\equiv  &
\left< R | J^\mu | N \right>.
\label{eqJNR0}
\ea
Next, we present the general definition of 
the helicity amplitudes which are valid 
for any final state resonance with spin $1/2$ or $3/2$.
Afterwards, we consider in particular the $\gamma^\ast N \to N (1520)$ 
and $\gamma^\ast N \to N (1535)$ transitions, 
and present the explicit expressions  
for the current and transition form factors.

Along this work we use a common notation 
for the two transitions.
The meaning of the index $R$,
as in the function ${\cal I}_R$ discussed previously,
depends on the transition under study.
We use also 
\ba
\tau= \frac{Q^2}{(M_R+M_N)^2},
\label{eqTau}
\ea
for both transitions.

\subsection{Helicity amplitudes}

The electromagnetic transition
$\gamma^\ast N \to R$, 
where $R$ is a state with 
angular momentum $J=\frac{1}{2}, \frac{3}{2}$
with positive or negative parity ($J^P=\frac{1}{2}^\pm, \frac{3}{2}^\pm$)
is characterized by the helicity amplitudes, 
functions of $Q^2$,
and defined at the $R$ rest frame by~\cite{Aznauryan12a}:
\ba
A_{3/2}&=&
\sqrt{\frac{2\pi \alpha}{K}}
\left< R, S_z'=+\sfrac{3}{2} \right|
\varepsilon_+ \cdot J \left| N,S_z=+\sfrac{1}{2} \right>, 
\nonumber  \\
& &
\label{eqA32} \\
A_{1/2}
&= &\sqrt{\frac{2\pi \alpha}{K}}
\left< R, S_z'=+\sfrac{1}{2} \right|
\varepsilon_+ \cdot J \left| N,S_z=-\sfrac{1}{2} \right>, 
\nonumber \\
\label{eqA12} \\
S_{1/2}&=&
\sqrt{\frac{2\pi \alpha}{K}}
\left< R, S_z'=+\sfrac{1}{2} \right|
\varepsilon_0 \cdot J \left| N,S_z=+\sfrac{1}{2} \right>\frac{|{\bf q}|}{Q}.
\nonumber \\
& &
\label{eqS12}
\ea
In the previous equations
 $S_z'$ ($S_z$) is the final (initial) 
spin projection,
${\bf q}$ is the photon three-momentum in the $R$ rest frame,
$Q=\sqrt{Q^2}$, $\varepsilon_\lambda^\mu$ ($\lambda=0,\pm 1$) is
the photon polarization vector,  
 $\alpha \simeq 1/137$ 
is the fine-structure constant  and 
$K= \frac{M_R^2-M^2}{2M_R}$.
The amplitude $A_{3/2}$ is defined only 
for $J=\frac{3}{2}$ resonances.

At the $R$ rest frame the magnitude of the photon three-momentum is
$|{\bf q}|$, and reads
\be
|{\bf q}|= \frac{\sqrt{Q_+^2Q_-^2}}{2M_R},
\label{eqq2}
\ee
where $Q_\pm^2= (M_R \pm M_N)^2 + Q^2$.
Note that when $Q^2=0$, one has $K =|{\bf q}|= \frac{M_R^2-M_N^2}{2M_R}$,
as mentioned above.

\subsection{$\gamma^\ast N \to N (1520)$ transition}

Because $N(1520)$ is a $J^P= \frac{3}{2}^-$ state,
the $\gamma^\ast N \to N(1520)$ 
transition current  
can be represented as~\cite{Aznauryan12a,D13}
\ba
J^\mu_{NR} =
\bar u_\beta(P_+) \Gamma^{\beta \mu} u (P_-),
\label{eqJNR1}
\ea
where $u_\beta$, $u$ are, respectively,
the Rarita-Schwinger and Dirac spinors. 
The operator $\Gamma^{\beta \mu}$ 
has the general Lorentz structure
\ba
\Gamma^{\beta \mu}=
G_1 q^\beta \gamma^\mu + 
G_2 q^\beta P^\mu + 
G_3 q^\beta q^\mu -
G_4 g^{\beta \mu},
\label{eqJ12}
\ea
where  
$P= \sfrac{1}{2}(P_+ + P_-)$.
The functions
$G_i$  ($i=1,..,4$) 
are form factor functions that depend on $Q^2$,
but only three of them are independent.
From current conservation~\cite{D13,Siegert2} we conclude that 
\ba
G_4=(M_R-M_N) G_1 + \frac{1}{2}(M_R^2-M_N^2)G_2 -Q^2 G_3.
\nonumber \\
\label{eqG4}
\ea
Another useful combination of the form factors $G_i$ 
\mbox{($i=1,2,3$)} is
 \ba
g_C&=& 4M_R G_1 + (3 M_R^2+M_N^2+Q^2)G_2 \nonumber \\
& & + 
2 (M_R^2-M_N^2-Q^2) G_3.
\label{eqGCsmall}
\ea

Using the previous form factors we can 
express the $\gamma^\ast N \to N (1520)$ 
helicity amplitudes defined by Eqs.~(\ref{eqA32})-(\ref{eqS12}) 
as~\cite{Aznauryan12a,D13}
\ba
\hspace{-.7cm}
& &
A_{1/2}=2  {\cal A}_R
\left\{
G_4 -\left[
(M_R-M_N)^2 +Q^2\right]
 \frac{G_1}{M_R} \right\}, \label{eqA12_D13} \\
\hspace{-.7cm}
& &
A_{3/2}=2 \sqrt{3} {\cal A}_R
G_4,  
\label{eqA32_D13} \\
\hspace{-.7cm}
& &
S_{1/2}= - \frac{1}{\sqrt{2}}
\frac{|{\bf q}|}{M_R} {\cal A}_R \,
g_C,  
\label{eqS12_D13}
\ea
where
${\cal A}_R= \frac{e}{4} \sqrt{\frac{Q_+^2}{6 M_N M_R K}}$.

For a discussion about the convenience of the combination of form factors 
$G_1$, $G_4$, $g_C$,  see Refs.~\cite{D13}.

An alternative representation of the 
$\gamma^\ast N \to N (1520)$ structure is the 
 so-called electromagnetic multipole form factors: 
the magnetic dipole ($G_M$), and the 
electric ($G_E$) and Coulomb ($G_C$) quadrupoles.
Those form factors can be represented as~\cite{Aznauryan12a,D13} 
\ba
\hspace{-.7cm}
G_M &=& -
{\cal R} \left[ (M_R-M_N)^2 +Q^2 \right] \frac{G_1}{M_R}, 
\label{eqGM}\\
\hspace{-.7cm}
G_E 
&= &-
{\cal R} \left\{ 4 G_4 - 
\left[ (M_R-M_N)^2 +Q^2 \right] \frac{G_1}{M_R} \right\}, 
\label{eqGE} \\
\hspace{-.7cm}
G_C &=&   - {\cal R} g_C,
\label{eqGC}
\ea
where 
$ {\cal R}=  \frac{1}{\sqrt{6}}\frac{M_N}{M_R-M_N}$.

\subsection{$\gamma^\ast N \to N (1535)$ transition}

We consider now the resonance 
$N(1535)$ which is a $J^P= \frac{1}{2}^-$ state.
The $\gamma^\ast N \to N (1535)$ transition current
can be represented as~\cite{Aznauryan12a,S11,S11b,S11c} 
\ba
J_{NR}^\mu=
\bar u_R 
\left[
F_1^\ast
\left(\gamma^\mu -\frac{\not \! q q^\mu}{q^2}\right) 
+ 
F_2^\ast
\frac{i \sigma^{\mu \nu} q_\nu}{M_R+ M_N} 
\right]
\gamma_5 u, \nonumber \\
\label{eqJS}
\ea
where 
$F_i^\ast$ ($i=1,2$) define the transition form factors
and 
$u_R$, $u$ are Dirac spinors associated
with the $R$ and the nucleon states, respectively.
The analytic properties
of the current (\ref{eqJS}) imply 
that $F_1^\ast(0)=0$ \cite{S11}.

The helicity amplitudes can be expressed in terms 
of the form factors using~\cite{Aznauryan12a,S11,Note,Siegert1}: 
\ba
A_{1/2} &=&  
2 {\cal A}_R  
\left[F_1^\ast + \frac{M_R-M_N}{M_R+M_N} F_2^\ast
\right]
\label{eqA12X}
\\
\hspace{-1cm}
S_{1/2} &=& 
- \sqrt{2} {\cal A}_R (M_R+M_N) \frac{|{\bf q}|}{Q^2} 
\nonumber \\
& &
\times 
\left[\frac{M_R-M_N}{M_R+M_N} F_1^\ast - \tau F_2^\ast
\right],
\label{eqS12X}
\ea
where
${\cal A}_R= \frac{e}{4} \sqrt{\frac{Q_+^2}{ M_N M_R K}}$.

We discuss next the results of 
the covariant spectator quark model for the 
transitions under discussion.


\section{Covariant spectator quark model}
\label{secCSQM}

The covariant spectator quark model
is derived from the formalism of
the covariant spectator theory~\cite{Gross}.
In the model, a baryon $B$ is described as a
three-constituent-quark system, where one quark is free to interact
with the electromagnetic fields and the other quarks are on-mass-shell.
Integrating over the on-mass-shell momenta,
one can represent the quark pair as an on-mass-shell
diquark with effective mass $m_D$,
and the baryon as a
quark-diquark system~\cite{NSTAR,Nucleon,Omega,Nucleon2}.
The structure of the baryon is then expressed by a
transition vertex between the three-quark bound state
and a quark-diquark state, that describes
effectively the confinement~\cite{Nucleon,Omega}.

The baryon wave function $\Psi_B(P,k)$ is 
derived from the transition vertex
as a function of the baryon momentum $P$ and
the diquark momentum $k$,
taking into account the properties of the baryon $B$,
such as the spin and flavor.
The wave functions are not determined by 
a  dynamical equation but are instead 
built from the baryon internal symmetries,
with the shape determined directly by
experimental or lattice data for 
some ground state systems~\cite{NSTAR,Nucleon,NDelta,LatticeD}.
The wave functions of the nucleon, $N(1520)$
and $N(1535)$ are discussed in Refs.~\cite{Nucleon,D13,S11}.

The covariant spectator
quark model was already applied to the 
nucleon~\cite{Nucleon,Nucleon2,NucleonDIS,FixedAxis,AxialFF},
several nucleon
resonances~\cite{N1710,Roper,SQTM,D13,S11}, 
$\Delta$ 
resonances~\cite{Deformation,NDelta,LatticeD,Delta,Lattice,Delta1600,SQTM}, 
and other transitions 
between baryon states~\cite{Omega,OctetFF,Octet2Decuplet,Omega2,S11c}.

When the baryon wave functions are represented
in terms of the single quark and
quark-pair states, one can write
the transition current
in a relativistic impulse approximation as~\cite{Nucleon,Omega,Nucleon2}
\ba
J_{NR}^\mu 
= 3 \sum_{\Gamma} \int_k
\bar \Psi_R (P_+,k) j_q^\mu \Psi_N (P_-,k),
\ea
where $j_q^\mu$ is the quark current operator 
and $\Gamma$ labels the scalar diquark
and vectorial diquark (projections $\Lambda=0,\pm$)
polarizations.
The factor 3 takes account of the contributions
of all the quark-pairs by symmetry.
The integration symbol represents the
covariant integration for the diquark on-shell state
$\int_k \equiv \int \frac{d^3 {\bf k}}{(2\pi)^3(2E_D)}$,
with $E_D=\sqrt{m_D^2 + {\bf k}^2}$.

The quark current operator can be written in
terms of the Dirac ($j_1$) and Pauli ($j_2$) quark
form factors~\cite{Nucleon,Omega}:
\ba
j_q^\mu = j_1 
\left(\gamma^\mu - \frac{\not \! q q^\mu}{q^2}
\right) + j_2 \frac{i \sigma^{\mu \nu} q_\nu}{2M_N}.
\ea
The inclusion of the term $- \frac{\not q q^\mu}{q^2}$
associated with the Dirac component
in inelastic reactions 
is equivalent to the Landau prescription
for the current $J^\mu$~\cite{Kelly98,Batiz98,Gilman02}.
The term restores current conservation,
but does not affect the results for the observables~\cite{Kelly98}.

In the $SU(2)$-flavor sector we can decompose ($i=1,2$)
\ba
j_i = \frac{1}{6} f_{i+} + \frac{1}{2} f_{i-} \tau_3. 
\ea
where $f_{i\pm}(Q^2)$ are quark electromagnetic form factors,
and normalized according to $f_{1\pm}(0)=1$
and $f_{2\pm}(0)= \kappa_\pm$ 
(quark isoscalar/isovector anomalous magnetic moment).

The quark electromagnetic form factors are 
written in  terms of a vector
meson dominance parametrization
that simulates effectively the
constituent quark internal structure
due to the interactions with gluons
and quark-antiquark polarization effects~\cite{Nucleon}.
The quark electromagnetic current was calibrated previously
by the nucleon and decuplet baryon data~\cite{Nucleon,Omega},
and was also tested in the lattice regime 
for the nucleon elastic reaction
as well as for the $\gamma^\ast N \to \Delta$
transition~\cite{OctetFF,Omega,Lattice,LatticeD}.
Details can be found in Refs.~\cite{Nucleon,Omega,Roper,N1710}.

In the calculation of the transition current 
it is convenient to define 
the symmetric ($S$) and antisymmetric ($A$) 
projections of the isospin states ($i=1,2$)
\ba
& &
j_i^S =\frac{1}{6}f_{i+} + \frac{1}{2}f_{i-} \tau_3 , 
\label{eqJiS}
\\
& &
j_i^A= \frac{1}{6}f_{i+} - \frac{1}{6}f_{i-} \tau_3.
\label{eqJiA}
\ea

The normalization of the states is imposed 
for $B=N,R$, through the condition~\cite{S11,D13}
\ba
\sum_\Gamma \int_k \bar \Psi_B (\bar P,k) 
(3 j_1) \gamma^0 \Psi_B(\bar P,k)=  e_N,
\label{eqNorm1}
\ea
where $\bar P= (M_B,0,0,0)$ is the momentum 
at the rest frame, $e_N= \frac{1}{2}(1 + \tau_3) $
is the nucleon charge.
In the previous equation $(3 j_1) \gamma^0$ 
is the quark charge operator, with $j_1=j_1(0)$.
Assumed in Eq.~(\ref{eqNorm1}) is 
the normalization of the radial wave function 
$\int_k  |\psi_B (\bar P,k)|^2 = 1$.

The radial wave functions $\psi_B$ ($B=N,R$) are represented 
in terms of the dimensionless variable \cite{Nucleon}
\ba
\chi= \frac{(M_B-m_D)^2-(P-k)^2}{M_B m_D},
\ea
as
\ba
\psi_B (P,k) =
\frac{N_0}{m_D( \beta_1 + \chi)(\beta_2 + \chi)},
\label{eqPsiR}
\ea
where $N_0$ is a normalization constant and 
the parameters 
$\beta_1 =0.049$ and $\beta_2=0.717$ 
were determined by the model for 
the nucleon with a fit to the nucleon 
electromagnetic form factor data~\cite{Nucleon}.
The relative sign of $N_0$ for the resonances 
$N(1520)$, $N(1535)$ is determined by the sign 
of the transition form factors~\cite{S11,D13}. 
With the inclusion of the factor  $1/m_D$
in the definition of the radial wave function (\ref{eqPsiR}),
the diquark mass dependence scales
out of in the integration ($k \to k/m_D$)
and the form factors became independent 
of $m_D$~\cite{Nucleon,Omega}.

The orthogonality condition between the nucleon and $R$ wave functions, 
now generalized with the effect of the isospin, is~\cite{D13,S11}:
\ba
\sum_\Gamma \int_k \bar \Psi_R (\bar P_+,k) 
(3 j_1) \gamma^0 \Psi_N(\bar P_-,k)= 0.
\label{eqOrthRel3}
\ea
From the previous expression we can derive 
the orthogonality condition for the radial wave functions, 
given by Eq.~(\ref{eqOrthRel2})~\cite{D13,S11}.

For the calculation of the transition 
form factors it is convenient to 
define the overlap integral 
between the radial wave functions ${\cal I}_R(Q^2)$ as
\ba
{\cal I}_R (Q^2) =
\int_k \frac{k_z}{|{\bf k}|} \psi_R(P_+,k) 
\psi_N( P_-,k),
\label{eqIR2}
\ea
The previous integral is calculated 
at the $R$ rest frame, where the integrate function is simplified.
The function ${\cal I}_R (Q^2)$ defines, however, 
an invariant integral that can be calculated in any frame.
The discussion associated with the general form 
of the integral 
can be found in Refs.~\cite{D13,S11}.
In the limit $Q^2=0$, we recover the form of ${\cal I}_R(0)$ 
presented in Eq.~(\ref{eqOrthRel2}).
As discussed in Sec.~\ref{secOrthogonality}, 
one has ${\cal I}_R (0) =0$, 
when $\psi_R$ is defined as $\psi_N$
($\psi_R \equiv \psi_N$) and $M_R=M_N$.

\subsection{$\gamma^\ast N \to N(1520)$ form factors}

The expressions for the $\gamma^\ast N \to N(1520)$ transition form factors
are~\cite{D13}
\ba
G_1&=& -\frac{3}{2\sqrt{2} |{\bf q}|}  \nonumber \\
& & \times
\left[
 \left( j_1^A + \frac{1}{3} j_1^S\right)+ 
\frac{M_R+M_N}{2M_N}\left( j_2^A + \frac{1}{3} j_2^S\right) 
\right] {\cal I}_R, \nonumber \\
& & \label{eqG1}\\
G_2 &=& 
 \frac{3}{2\sqrt{2} M_N |{\bf q}|},   
\nonumber \\
& & \times
\left[
j_2^A 
+ \frac{1}{3}  \frac{1- 3 \tau}{1 + \tau}   j_2^S
+ 
\frac{4}{3} \frac{2 M_N}{M_R +M_N} \frac{1}{1 + \tau} j_1^S 
\right] {\cal I}_R, \nonumber \\
& & 
\label{eqG2}
 \\
G_3 &=& -\frac{3}{2\sqrt{2} |{\bf q}|} 
\frac{M_R-M_N}{Q^2}  
\nonumber \\
& &
\times \left[
j_1^A + \frac{1}{3}\frac{\tau-3}{1+ \tau} j_1^S 
+ \frac{4}{3} \frac{M_R +M_N}{2M_N} \frac{\tau}{1+ \tau} j_2^S
\right] {\cal I}_R, \nonumber \\
\label{eqG3}
\ea 
where $\tau$ is defined by Eq.~(\ref{eqTau}). 

Comparative with Refs.~\cite{D13} we take 
the limit where the mixture angle $\theta_D$ 
is given by $\cos \theta_D \simeq 1$
(in most of the models $\cos \theta_D \simeq 0.99$ \cite{Burkert03}).

When we calculate $G_4$ using (\ref{eqG4}), 
we obtain 
\ba
G_4=0.
\ea
Thus, in the covariant spectator quark model one has,
according to Eqs.~(\ref{eqA32_D13}), 
(\ref{eqGM}) and (\ref{eqGE}):
\mbox{$A_{3/2} \equiv 0$} and $G_E \equiv - G_M$~\cite{D13}.

For the following discussion we note that the form factors $G_i$ ($i=1,2,3$) 
given by Eqs.~(\ref{eqG1})-(\ref{eqG3})
are proportional to the factor $\frac{{\cal I}_R(Q^2)}{|{\bf q}|}$.

The results (\ref{eqG1})-(\ref{eqG3}) were derived in Refs.~\cite{D13}.
In that work $\psi_R$ was parametrized in 
order to describe the large $Q^2$ data (small meson cloud effects)
and the orthogonality condition, ${\cal I}_R(0)=0$.
As a consequence the valence quark contributions 
for the form factors $G_M$, $G_E$ and the amplitude $A_{1/2}$ 
vanishes at $Q^2=0$.
This feature changes in the semirelativistic approximation, 
as we show later.

Another interesting property of the model is that the 
results for the form factors $G_i$ imply that $A_{3/2} =0$
for all values of $Q^2$, in contradiction 
with the available experimental data.
Our interpretation of this result is that
the main contribution for the amplitude  $A_{3/2} $
comes from the meson cloud effects.
This assumption is consistent 
with the results presented in the literature.
Most of the quark models 
predict only small contributions 
for the  amplitude $A_{3/2}$ 
(about 1/3 of the empirical 
data)~\cite{Warns90,Aiello98,Merten02,Santopinto12,Ronniger13},
although there are exceptions~\cite{Capstick95}.
A more detailed discussion can be found in Refs.~\cite{D13,SQTM}.
Estimates of the meson cloud contributions 
from the EBAC coupled-channel reaction model 
also support the idea that 
the meson cloud is the dominant effect in $A_{3/2} $~\cite{EBAC}.

\subsection{$\gamma^\ast N \to N(1535)$ form factors}

The expressions for the 
$\gamma^\ast N \to N(1535)$ 
transition form factors are~\cite{D13}
\ba
& &
F_1^\ast(Q^2)= \frac{1}{2}(3j_1^S+ j_1^A)  {\cal I}_R,  
\label{eqF1}\\
& &
F_2^\ast(Q^2)= -\frac{1}{2}(3j_2^S- j_2^A) \frac{M_R+M_N}{2M_N} {\cal I}_R. 
\label{eqF2}
\ea

In Ref.~\cite{S11}, we presented a model with ${\cal I}_R(0) \ne 0$.
The consequence of ${\cal I}_R(0) \ne 0$
is that the nucleon and the resonance $N(1535)$
are not orthogonal.
The results presented in Ref.~\cite{S11} 
were based on an approximated orthogonality,
and are valid only for large $Q^2$ 
($Q^2 \gg K^2 \simeq  0.2$ GeV$^2$).
In the present work, within the semirelativistic approximation, 
the orthogonality is exact.


\section{Results in the Semirelativistic approximation} 
\label{secResults}


We present now the results of the 
semirelativistic approximation
for the $\gamma^\ast N \to N(1520)$ and 
the $\gamma^\ast N \to N(1535)$ transition form factors
and respective helicity amplitudes.

The numerical results are compared 
to the data from CLAS single pion production
\cite{CLAS1}, CLAS double pion production \cite{CLAS2,Mokeev16},
MAID \cite{MAID1,MAID2} and
Particle Data Group (PDG) ($Q^2=0$)~\cite{PDG}.
For the $\gamma^\ast N \to N(1535)$ transition
we also present  results from 
JLab/Hall C~\cite{Dalton09} for the amplitude $A_{1/2}$.

\subsection{$\gamma^\ast N \to N(1520)$ transition}

The elementary form factors $G_i$ ($i=1,2,3$)
for the  $\gamma^\ast N \to N(1520)$ transition, 
determined by the covariant spectator quark 
model, are expressed by Eqs.~(\ref{eqG1})-(\ref{eqG3}).
Using those expressions  for $G_i$ 
we can evaluate the helicity amplitudes
$A_{1/2}$, $A_{3/2}$, $S_{1/2}$ 
and the multipole form factors $G_M$, $G_E$ and $G_C$.

In the semirelativistic approximation
we evaluate $G_1$, $G_2$, $G_3$, in the limit 
$M_R = M_N$, 
using 
\ba
\frac{M_R+ M_N}{2 M_N} \to 1 , \\
|{\bf q}| \to Q \sqrt{1+ \tau},
\ea
and replacing also $M_N \to M$ in $G_2$.
Special care is necessary for the function $G_3$,
since it includes a factor $(M_R-M_N)/Q^2$.
There is therefore the possibility 
of a singularity at $Q^2=0$.
This singularity is only apparent as we explain next.

We start noticing that $G_3$ appears only in the 
function $g_C$ given by Eq.~(\ref{eqGCsmall}),
which can be expressed, in the limit $M_R= M_N$, as
\ba
g_C= 4M G_1 + (4M^2+ Q^2)G_2 - 2 Q^2 G_3.
\ea
Since the factor $1/Q^2$ in $G_3$
is canceled by the factor $Q^2$, 
the limit $M_R= M_N$ can be performed, obtaining $Q^2 G_3  \to 0$.

Note also that in $G_4$, we can drop the term $Q^2 G_3$.
We then conclude that in the  
limit $M_R=M_N$ all terms in $G_4$ vanish [see Eq.~(\ref{eqG4})].

We recall, from the previous section, 
that, the form factors $G_i$ are proportional 
to $\frac{{\cal I}_R(Q^2)}{|{\bf q}|}$.
Since ${\cal I}_R(Q^2) \propto |{\bf q}|$ near $Q^2=0$, 
when $N$ and $R$ are defined by the same radial wave function~\cite{S11},
in the approximation $M_R=M_N$,  
we ensure the orthogonality between the 
nucleon and resonance states, ${\cal I}_R(0) =0$,
and obtain also finite results at $Q^2=0$ 
(${\cal I}_R (0) /|{\bf q}| \ne 0$).

To calculate the helicity amplitudes and 
the form factors $G_M$, $G_E$ and $G_C$
in the semirelativistic approximation,  
we use the relations (\ref{eqA12_D13})-(\ref{eqS12_D13}),  
(\ref{eqGM})-(\ref{eqGC}), respectively, 
including as input the functions $G_i$ ($i=1,2,3$), $G_4$, $g_C$ 
determined in the limit $M_R= M_N$.

The conversion between $G_1$, $G_4$, $g_C$
into helicity amplitudes and multipole form factors 
using coefficients dependent on the physical masses $M_R$ and $M_N$
is necessary,  because the helicity amplitudes 
and $G_M$, $G_E$ and $G_C$ are  strictly defined only
in the case $M_R \ne M_N$.
At the end we present also the results 
in the {\it extreme limit}, when we ignore 
all mass differences, 
except for the factors ${\cal A}_R$ or ${\cal R}$.
It is worth to mention that 
the extreme limit is just a theoretical 
exercise, since as it is discussed later, 
it changes the properties of the 
multipole form factors and 
helicity amplitudes near $Q^2=0$.

\subsubsection{Comparison with the data}

\begin{figure}[t]
\vspace{.3cm}
\centerline{
\mbox{
\includegraphics[width=2.8in]{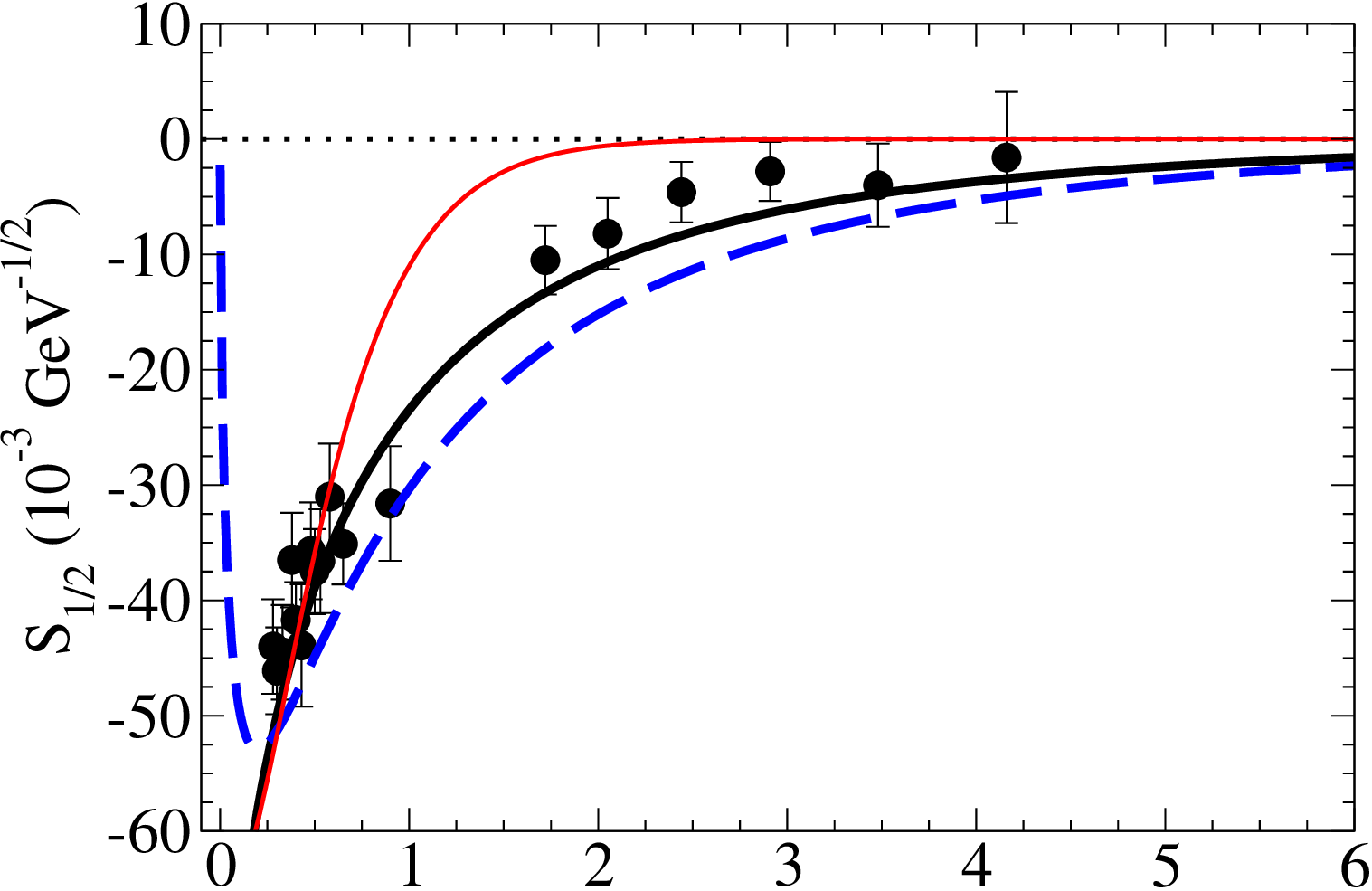}
}}
\centerline{
\vspace{.5cm} }
\centerline{
\mbox{
\includegraphics[width=2.8in]{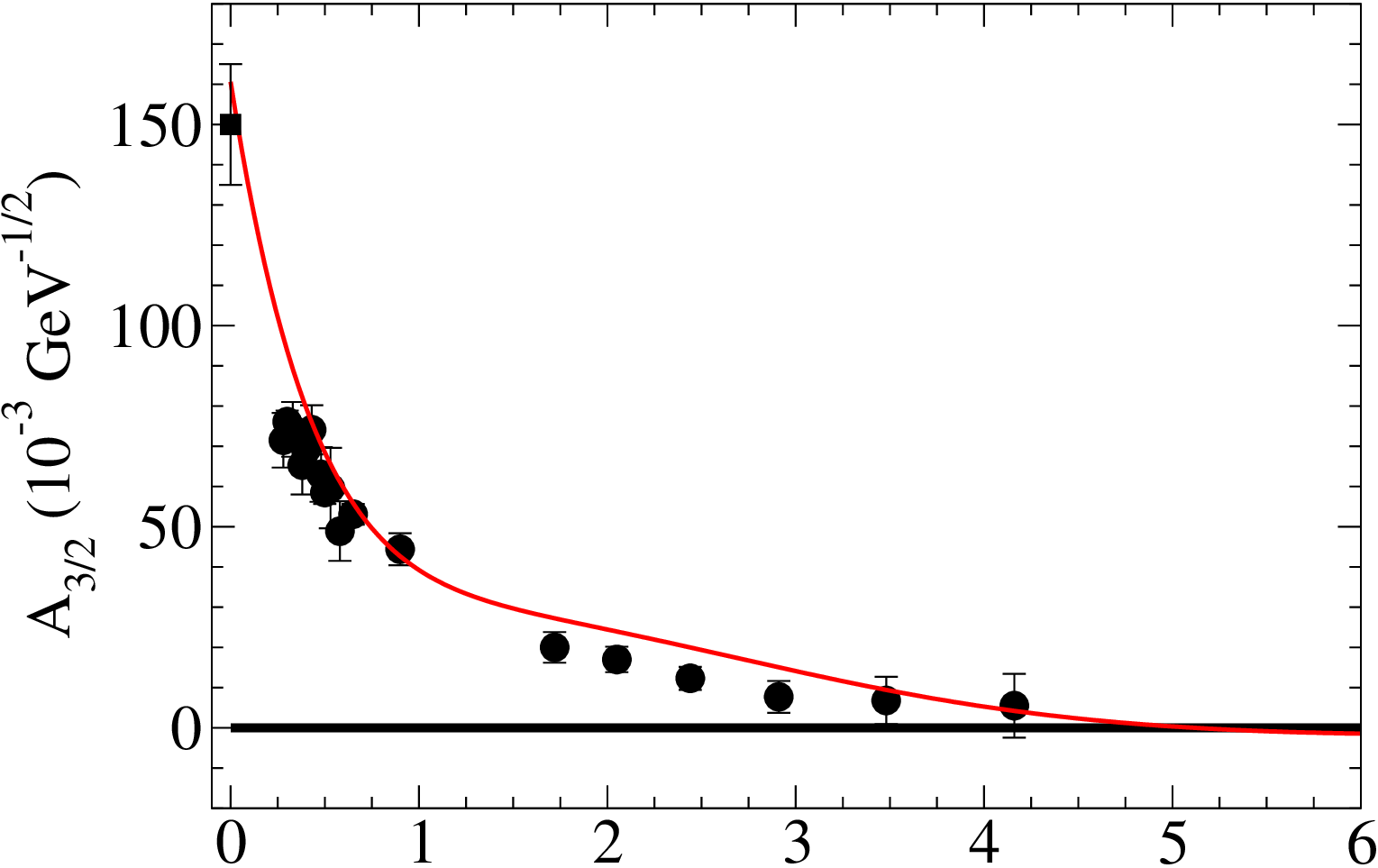}
}}
\centerline{
\vspace{.5cm} }
\centerline{
\mbox{
\includegraphics[width=2.8in]{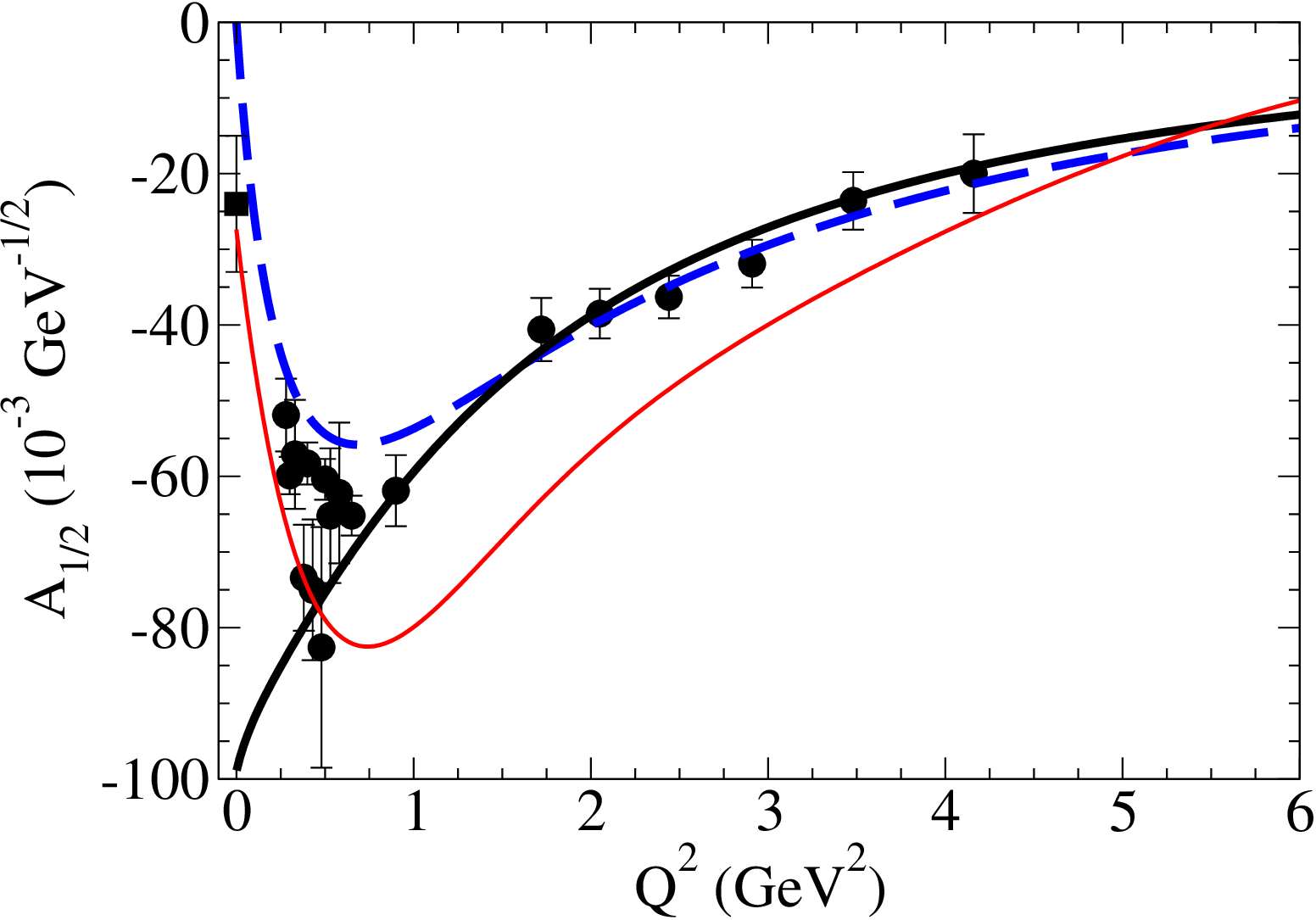}
}}
\caption{\footnotesize{
Results of the 
$\gamma^\ast N \to N(1520)$ helicity amplitudes
given by the semirelativistic approximation (thick solid line).
The semirelativistic approximation include only 
the effect of the valence quark core.
Data from  PDG \cite{PDG} (full squares) and  
CLAS \cite{CLAS1,CLAS2,Mokeev16} (full circles).
The thin solid line represent the fit to the MAID data \cite{MAID2}.
The {\it extreme limit} is represented by the dashed line.
}}
\label{figD13amps}
\end{figure}

The results of the semirelativistic approach (thick solid line)
for the helicity amplitudes
are present in Fig.~\ref{figD13amps}.
The CLAS data~\cite{CLAS1,CLAS2,Mokeev16} are represented 
by the full circles. 
For a cleaner comparison, we replace 
the MAID data by the MAID parametrization 
of the data \cite{MAID2} (thin solid line).
Notice the deviation between the MAID and the CLAS data 
for the amplitudes $A_{1/2}$ and $S_{1/2}$.
It is interesting to see in the figure that, 
the semirelativistic approximation  
describes very well the 
CLAS amplitudes $A_{1/2}$ and $S_{1/2}$ for $Q^2 > 1$ GeV$^2$.
As for the $A_{3/2}$, as discussed already, 
the model predicts $A_{3/2}  \equiv 0$.

\begin{figure}[t]
\vspace{.3cm}
\centerline{
\mbox{
\includegraphics[width=2.8in]{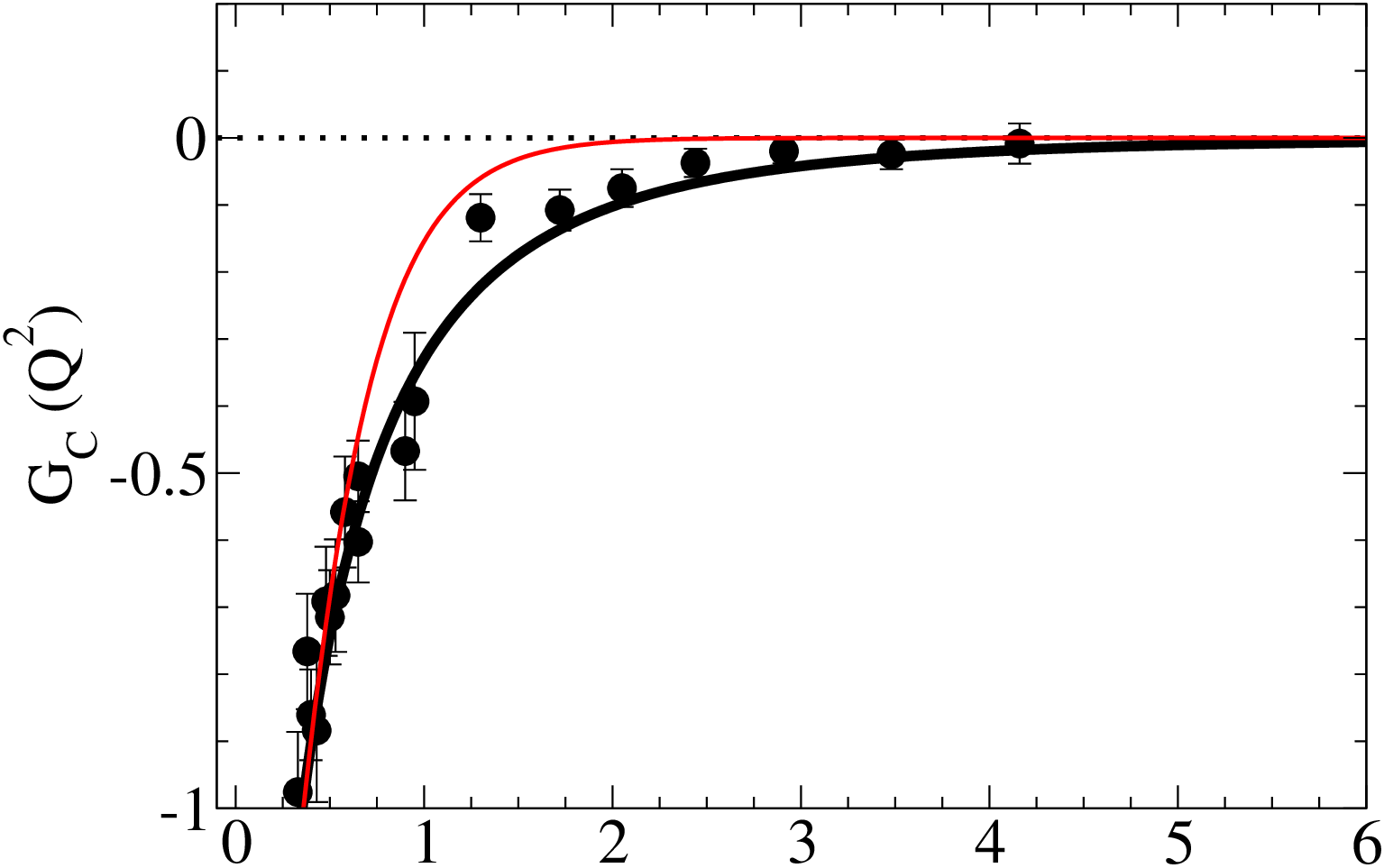}
}}
\centerline{
\vspace{.4cm} }
\centerline{
\mbox{
\includegraphics[width=2.8in]{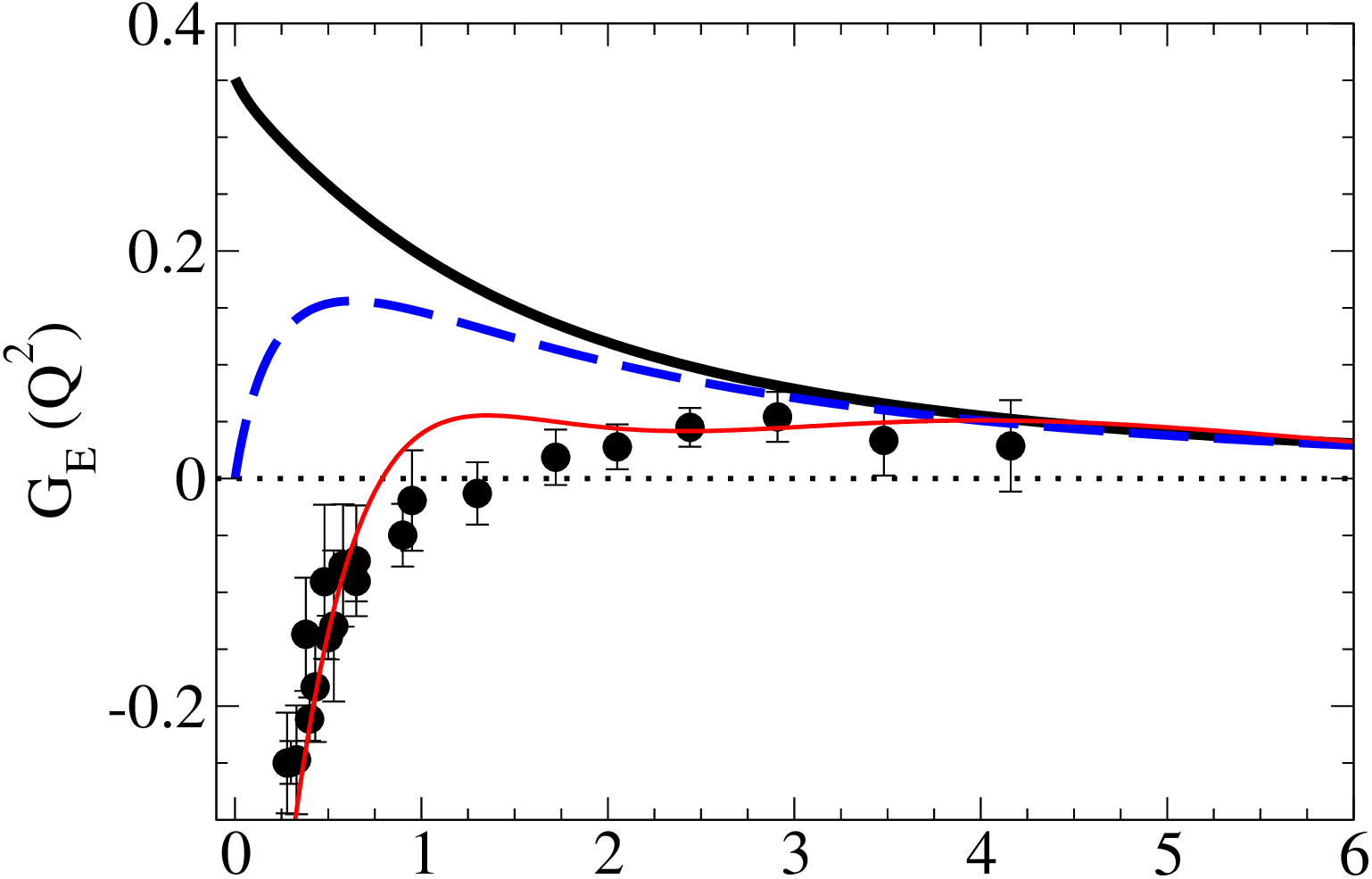}
}}
\centerline{
\vspace{.4cm} }
\centerline{
\mbox{
\includegraphics[width=2.8in]{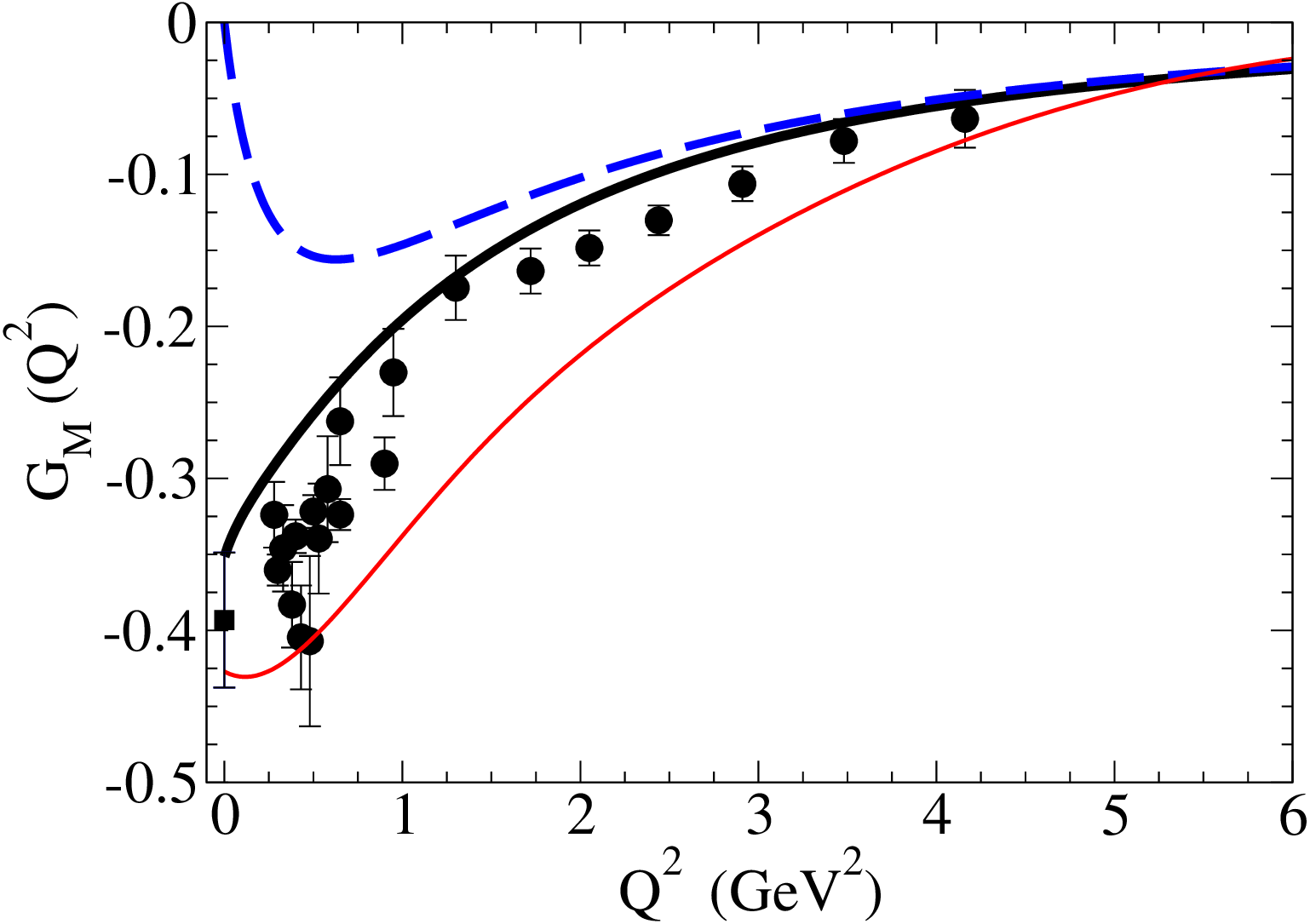}
}}
\caption{\footnotesize{
Results of the 
$\gamma^\ast N \to N(1520)$ form factors
given by the semirelativistic approximation (thick solid line).
The semirelativistic approximation include only 
the effect of the valence quark core.
Data from PDG \cite{PDG} (full squares) and 
CLAS \cite{CLAS1,CLAS2,Mokeev16} (full circles).
The thin solid line represent the fit to the MAID data \cite{MAID2}.
The {\it extreme limit} is represented by the dashed line.
}}
\label{figD13FF}
\end{figure}

The corresponding results for the 
form factors  $G_M$, $G_E$ and $G_C$ are presented
in Fig.~\ref{figD13FF}, with the same labeling.
The differences between the CLAS and MAID 
data are obvious for $G_M$ and $G_C$.
It is interesting to note 
in this case, that, although the 
semirelativistic approximation fails to  
describe the $G_E$ data at low $Q^2$,
it approaches the data for $Q^2 > 3$ GeV$^2$.

Overall it is remarkable the agreement 
between the model and the CLAS form factor data for 
intermediate and large $Q^2$.
Except for $A_{3/2}$, the comment is also valid  
for the helicity amplitudes.
We recall that the  large $Q^2$ behavior is a prediction 
of the model since no parameters are included for the resonance $R$.
The radial wave function associated with the resonance $R$
uses the parameters of the nucleon radial wave function
(same momentum distribution).

In both analysis, helicity amplitudes or multipole form factors,
the semirelativistic 
approach deviates from the CLAS data for small $Q^2$.  
Although our calculations are restricted 
to the limit $M_R=M_N$, we can still assume
that the main reason for the deviation at small $Q^2$ 
is due to the absence of the meson cloud effects in our formalism,
 since the meson cloud effects can be significant 
for some resonances at low $Q^2$.

In the graphs for $A_{1/2}$ and $G_M$ one can see 
that the semirelativistic approximation 
is very close to the data, particularly 
for $Q^2 > 1$ GeV$^2$.
We then conclude that, those functions 
are dominated by the valence quark effects 
(small meson cloud contributions).

Our results for $G_E$ are in
strong disagreement with the experimental data.
This result suggests that the form factor $G_E$
may have significant contributions from the meson cloud,  
in order to cover the gap between the model and the empirical data.
Recall that a similar effect was already observed  
for the amplitude $A_{3/2}$. 
The conclusion that $G_E$ is dominated by
meson cloud effects is one of the 
more important results of the present work.   
The results for $G_E$ and the connection 
with $A_{3/2}$ are discussed in more detail 
at the end of the section.


\subsubsection{Extreme limit}

In order to study in more detail 
the result of the approximation $M_R=M_N$,
we consider at last, the {\it extreme limit},
where we take also the $M_R=M_N$ limit
in the form factor coefficients of 
Eqs.~(\ref{eqA12_D13})-(\ref{eqS12_D13})
and  Eqs.~(\ref{eqGM})-(\ref{eqGC}).
The results are presented in Figs.~\ref{figD13amps}
and \ref{figD13FF} by the dashed line.
In that case we use $Q_-^2  = Q^2$,
and replace also $|{\bf q}| \to Q\sqrt{1 + \tau}$ in $S_{1/2}$.
For $G_C$ there is no difference between 
the semirelativistic approximation 
and the {\it extreme limit}.
As a consequence of the {\it extreme limit},
the functions $A_{1/2}$, $S_{1/2}$, $G_E$ and $G_M$ vanish at $Q^2=0$.
The form factor $G_C$ does not vanish 
at $Q^2=0$, because  the factor  $|{\bf q}|$
cancels the $Q \to 0$ dependence of $S_{1/2}$
(since $G_C \propto S_{1/2} /|{\bf q}|$).
One concludes then that the 
{\it extreme limit} modifies  
the behavior of the helicity amplitudes and 
multipole form factors at low $Q^2$,
particularly near $Q^2=0$,
and is in contradiction with the data 
(nonzero results for $A_{1/2}(0)$, $S_{1/2}(0)$, 
$G_E(0)$ and $G_M(0)$).  
For that reason the {\it extreme limit} 
should be seen as a theoretical exercise
that may differ from the physical case.
It is nevertheless interesting to note 
that the {\it extreme limit} is close to the CLAS data 
at low $Q^2$ for the amplitudes $A_{1/2}$ and  $S_{1/2}$.

\subsection{$\gamma^\ast N \to N(1535)$ transition}

We present now the results of the semirelativistic 
approximation for the  $\gamma^\ast N \to N(1535)$ transition.
We start with the discussion of the 
transition form factors, later we 
discuss the helicity amplitudes.

The available data for the $A_{1/2}$ 
and $S_{1/2}$ amplitudes cover the region $Q^2=0 -4.2$ GeV$^2$ 
\cite{PDG,CLAS1,MAID1,MAID2}.
The  large $Q^2$ data for $A_{1/2}$ come from 
Ref.~\cite{Dalton09} with $Q^2=5.8, 7.0$ GeV$^2$,
and were extracted under the assumption 
that the  $S_{1/2}$ contribution for 
the cross section is negligible.
Therefore, in the conversion from 
helicity amplitudes to transition form factors, we use  $S_{1/2}=0$.

In Ref.~\cite{S11b} it was suggested that 
in the region  $Q^2 > 2$ GeV$^2$
the amplitudes are related by 
$S_{1/2} =- \frac{\sqrt{1 + \tau}}{\sqrt{2}} \frac{M^2_R-M_N^2}{2 M_R Q} A_{1/2}$.
One can then use the relation to estimate 
the expected value for $S_{1/2}$ according to 
the values $A_{1/2}$ from Ref.~\cite{Dalton09} 
for large $Q^2$.
In the following we use the solid triangles 
for the original result ($S_{1/2}=0$)
and the empty triangles for the asymptotic estimate.


\begin{figure}[t]
\vspace{.3cm}
\centerline{
\mbox{
\includegraphics[width=2.8in]{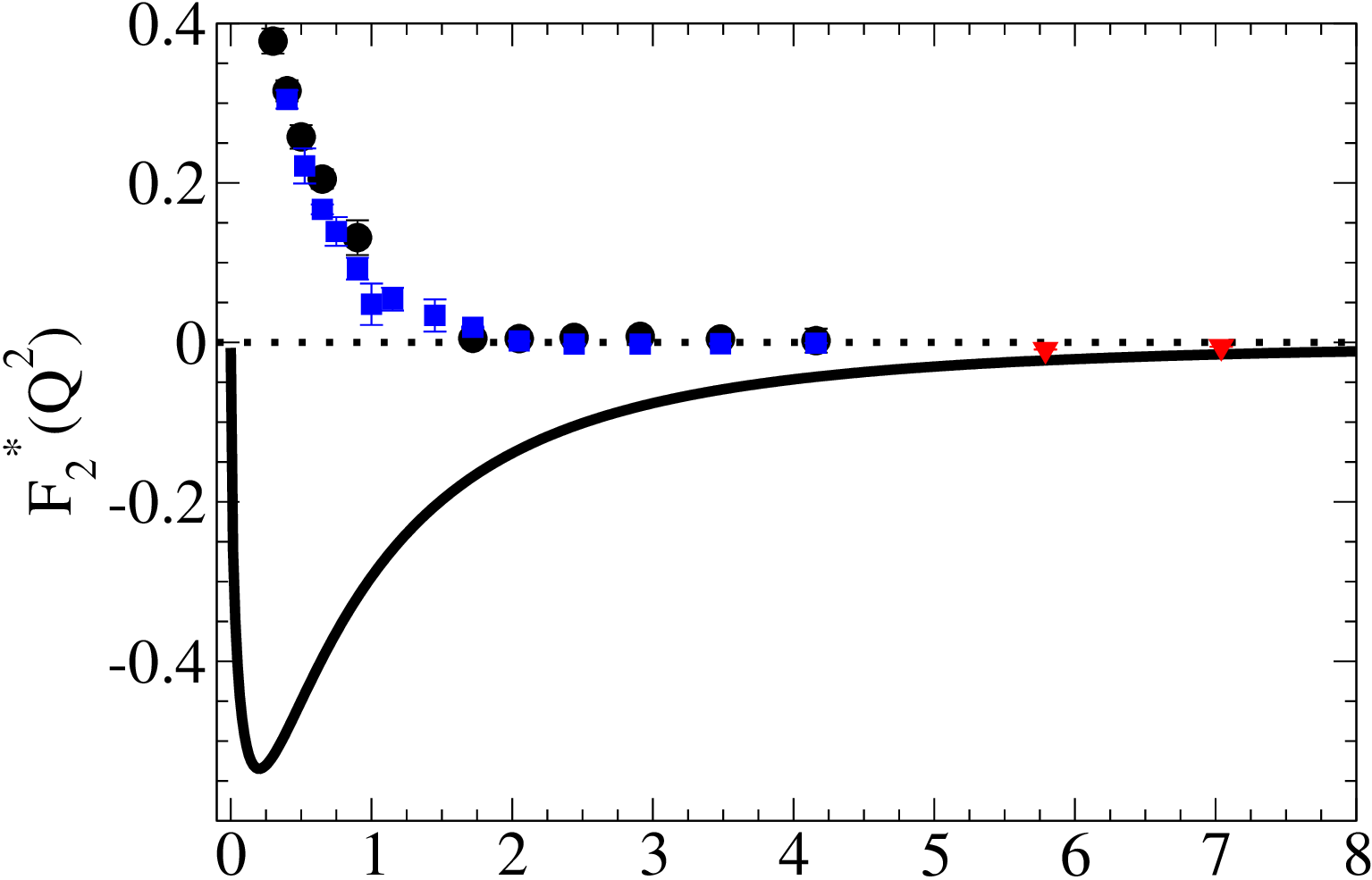}
}}
\centerline{
\vspace{.5cm} }
\centerline{
\mbox{
\includegraphics[width=2.8in]{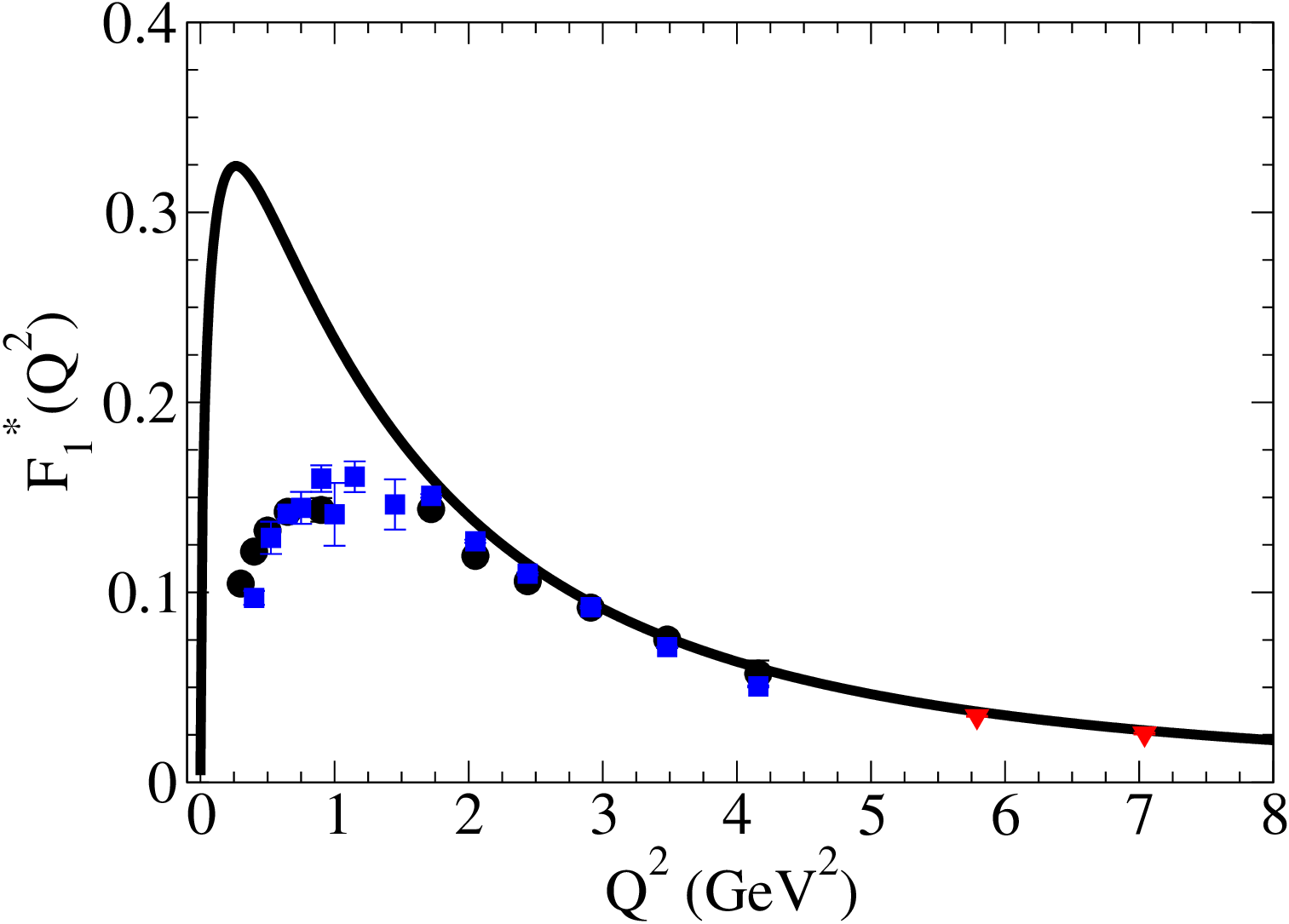}
}}
\caption{\footnotesize{
Results for the 
$\gamma^\ast N \to N(1535)$ transition form factors
given by the semirelativistic approximation (thick solid line).
The semirelativistic approximation include only 
the effect of the valence quark core.
Data from CLAS \cite{CLAS1} 
(full circles), MAID \cite{MAID1,MAID2} 
(full squares),
JLab/Hall C \cite{Dalton09} (triangles).  
Out of the range is the PDG result  $F_2^\ast(0) = 0.83  \pm   0.28$ \cite{PDG}.
}}
\label{figS11FF}
\end{figure}

\subsubsection{Form factors}

We start with the $\gamma^\ast N \to N(1535)$ 
results for the form factors $F_1^\ast$ and $F_2^\ast$.
In the calculation of the overlap integral ${\cal I}_R(Q^2)$,
we use the replacement of $M_R,M_N \to M$.
In the calculation of the form factors we 
consider in addition the replacement  
$\frac{M_R+ M_N}{2 M_N} \to 1$, in the expression for $F_2^\ast$.
In the semirelativistic approach, since 
$F_i^\ast(Q^2) \propto {\cal I}_R(Q^2)$ and ${\cal I}_R(0) =0$,
one has $F_1^\ast(0)=0$, and $F_2^\ast(0)=0$.
The first result is consistent with the data 
(by construction).
The second result is an approximation of our model, 
since the experimental value is $F_2^\ast(0) = 0.83  \pm   0.28$ \cite{PDG}.

The results for the form factors are presented in Fig.~\ref{figS11FF}
and are compared with the data from CLAS, MAID 
and JLab/Hall C \cite{CLAS1,MAID1,MAID2,Dalton09}
for $Q^2 > 0$.

In Fig.~\ref{figS11FF}, one can note for $F_1^\ast$,
the good agreement between the model (solid line)
and the data (CLAS and MAID) for $Q^2 > 2$ GeV$^2$.
As for $F_2^\ast$, we conclude as in the previous work~\cite{S11},
that the model predictions for $F_2^\ast$ 
are not in agreement with the data
(difference of sign between the model and the data).

Our interpretation of the results for $F_2^\ast$
is that the difference between the model and the data 
is due to the the meson cloud effects,
not included in our framework.
In that case we expect significant 
meson cloud contributions for $F_2^\ast$.
Our hypothesis is corroborated 
by explicit calculations of meson cloud 
effects based on the unitary chiral model, 
where the baryons states are 
represented by bare cores dressed by mesons~\cite{S11c,Jido08}.
As for $F_1^\ast$ the model describes very well 
the experimental data except for the region $Q^2 < 1.5$ GeV$^2$.
This result suggests that the missing effect in $F_1^\ast$ 
for small $Q^2$ may also be due to the meson cloud contributions.
For larger values of $Q^2$, the meson cloud effects 
are smaller and the form factor $F_1^\ast$ is dominated by 
valence quark effects, as expected.

\subsubsection{Helicity amplitudes}

In the case of the $\gamma^\ast N \to N(1535)$ transition 
there is no simple procedure to calculate the helicity amplitudes 
using our results in the semirelativistic approximation (limit $M_R=M_N$),
since in the calculation of the 
helicity amplitudes (\ref{eqA12X}) and (\ref{eqS12X})
the mass difference is crucial.
If we use $M_R=M_N$ in $A_{1/2}$, 
we suppress the contribution from $F_2^\ast$,
and $A_{1/2}$ is determined exclusively by $F_1^\ast$.
If we use $M_R=M_N$ in $S_{1/2}$, 
we remove the effect of  $F_1^\ast$,
the more relevant form factor.

Our first conclusion then is that 
the semirelativistic approximation 
to the $\gamma^\ast N \to N(1535)$ transition
is better for the form factors 
than for the helicity amplitudes.

To compare our estimates in the semirelativistic approach 
we need to consider additional simplifications.
We consider the two following cases:
\begin{itemize}
\item
{\bf Model A}  (or valence quark model)\\
It is 
defined by the semirelativistic approach 
to the form factors with no further constraints.
Since no explicit meson cloud effects are included, 
the model is expected to fail 
the description of the data at low $Q^2$.
It may happen, however, that the model is comparable 
with other estimates of the bare core.
\item
{\bf Model B} (or high $Q^2$ model)  \\
It is defined by the condition $F_2^\ast=0$, 
combined with the result of the model for $F_1^\ast$. 
Since the result $F_2^\ast=0$ holds only for high $Q^2$, 
the model is expected to be good only for large values of $Q^2$.
\end{itemize}

In both cases we use the original definition 
of amplitudes  (\ref{eqA12X}) and (\ref{eqS12X}).
In the analysis we discuss also 
the effect of the replacement 
 $|{\bf q}| \to Q \sqrt{1 + \tau}$ 
in the amplitude $S_{1/2}$ given by Eq.~(\ref{eqS12X}).
With the previous correction, 
$S_{1/2}$ became
\ba
S_{1/2}&=& 
\sqrt{2} {\cal A}_R (M_R+M_N) (1 + \tau) 
\nonumber \\
& &
\times 
\left[\frac{M_R-M_N}{M_R+M_N} \frac{F_1^\ast}{|{\bf q}|} - 
\tau \frac{F_2^\ast}{|{\bf q}|}
\right],
\label{eqS12t}
\ea
where $\frac{F_i^\ast}{|{\bf q}|}$ ($i=1,2$) are well defined 
functions at $Q^2=0$, as discussed already.
With the form (\ref{eqS12t}) the divergence in $1/Q^2$ 
of $S_{1/2}$ is avoided and $S_{1/2}(0)$ becomes finite.

\begin{figure}[t]
\vspace{.3cm}
\centerline{
\mbox{
\includegraphics[width=2.8in]{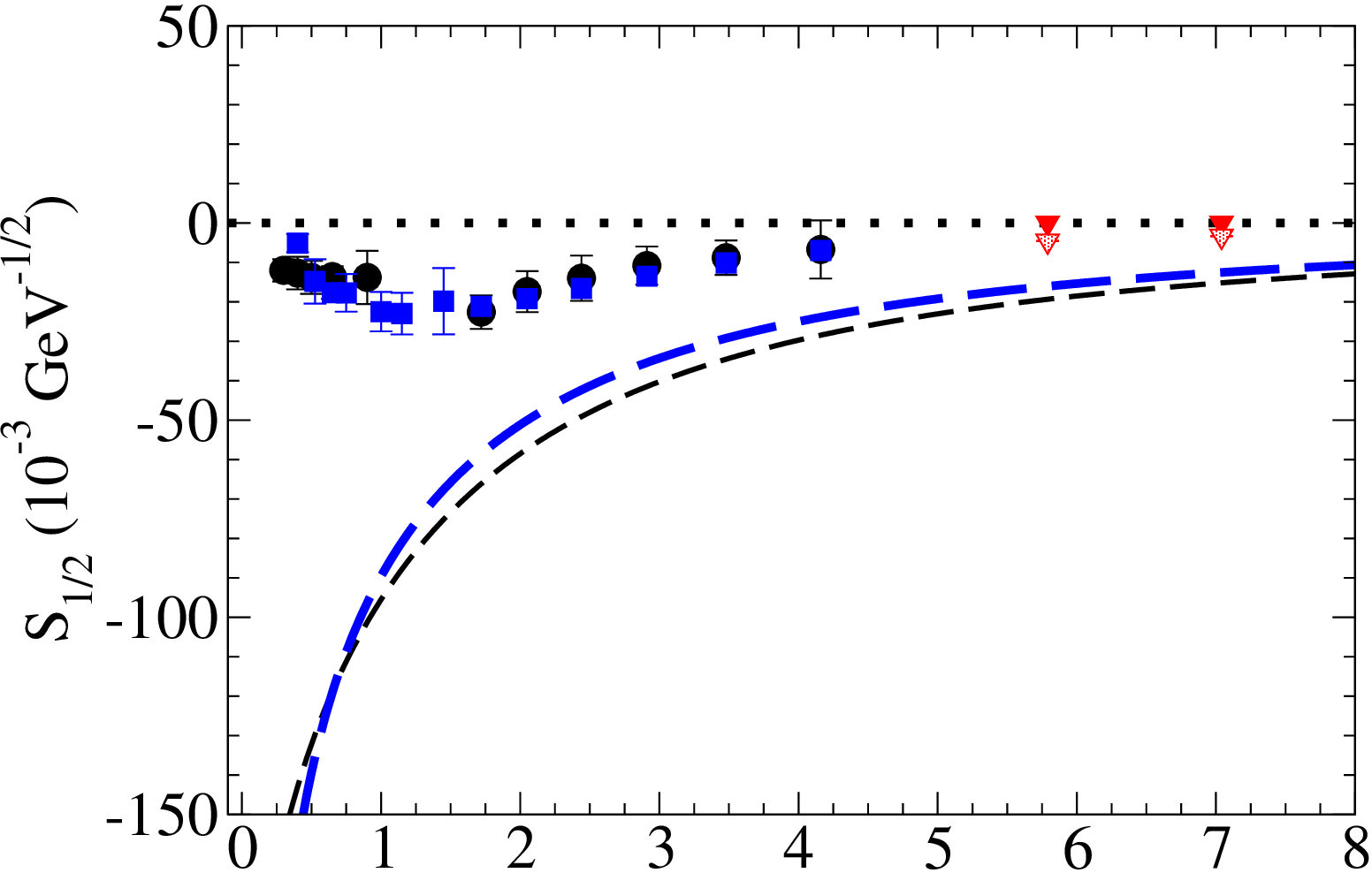}
}}
\centerline{
\vspace{.5cm} }
\centerline{
\mbox{
\includegraphics[width=2.8in]{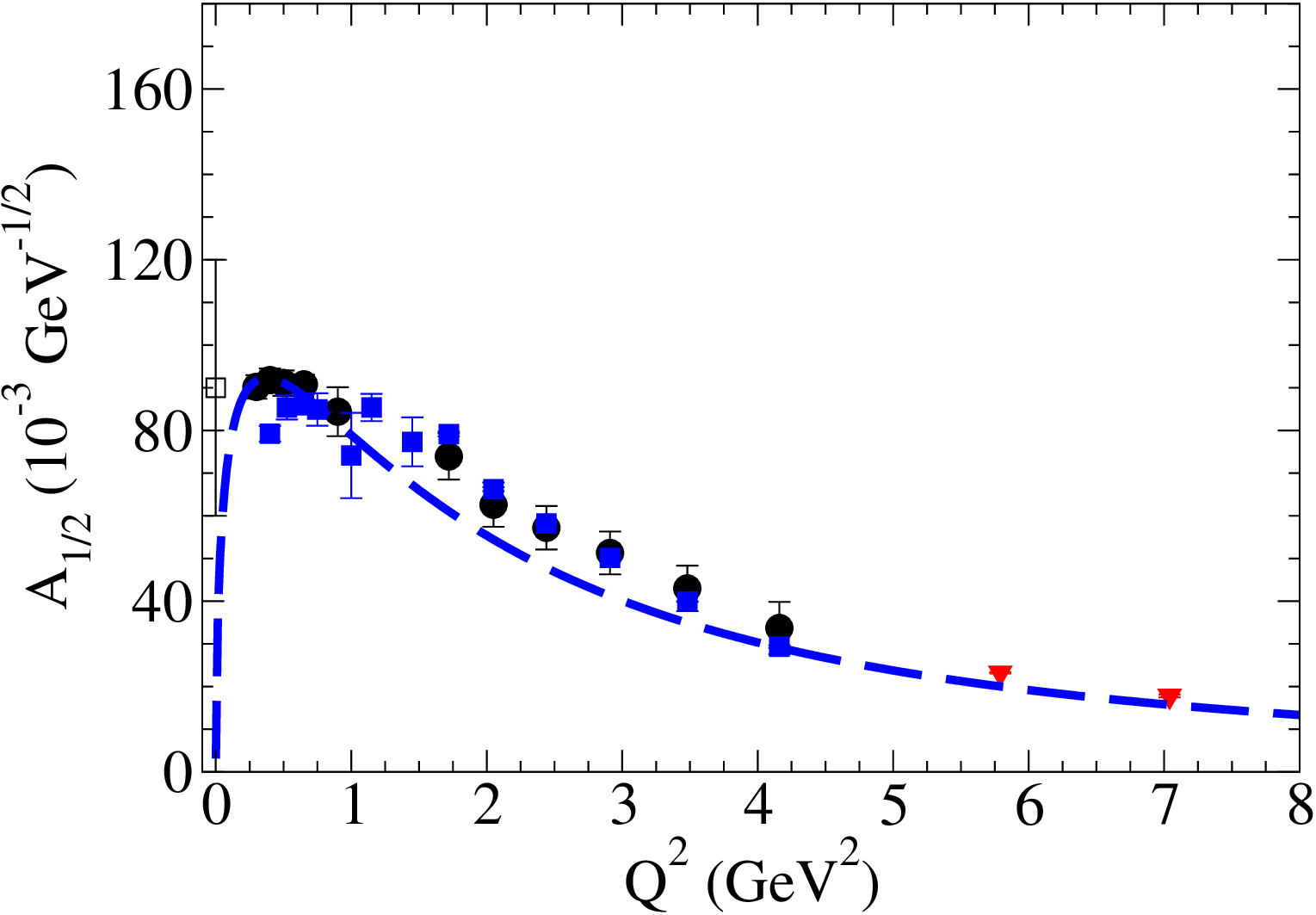}
}}
\caption{\footnotesize{
Results for the 
$\gamma^\ast N \to N(1535)$ helicity amplitudes given 
by the model A (thick dashed line). 
Model A is based in the valence quark effects 
(see description in the main text). 
The thin dashed line is the result of Eq.~(\ref{eqS12t}).
Data from PDG (empty square) \cite{PDG}, CLAS \cite{CLAS1} 
(full circles), MAID \cite{MAID1,MAID2} 
(full squares),
JLab/Hall C \cite{Dalton09} (triangles).  
}}
\label{figS11amps1}
\end{figure}

\begin{figure}[t]
\vspace{.3cm}
\centerline{
\mbox{
\includegraphics[width=2.8in]{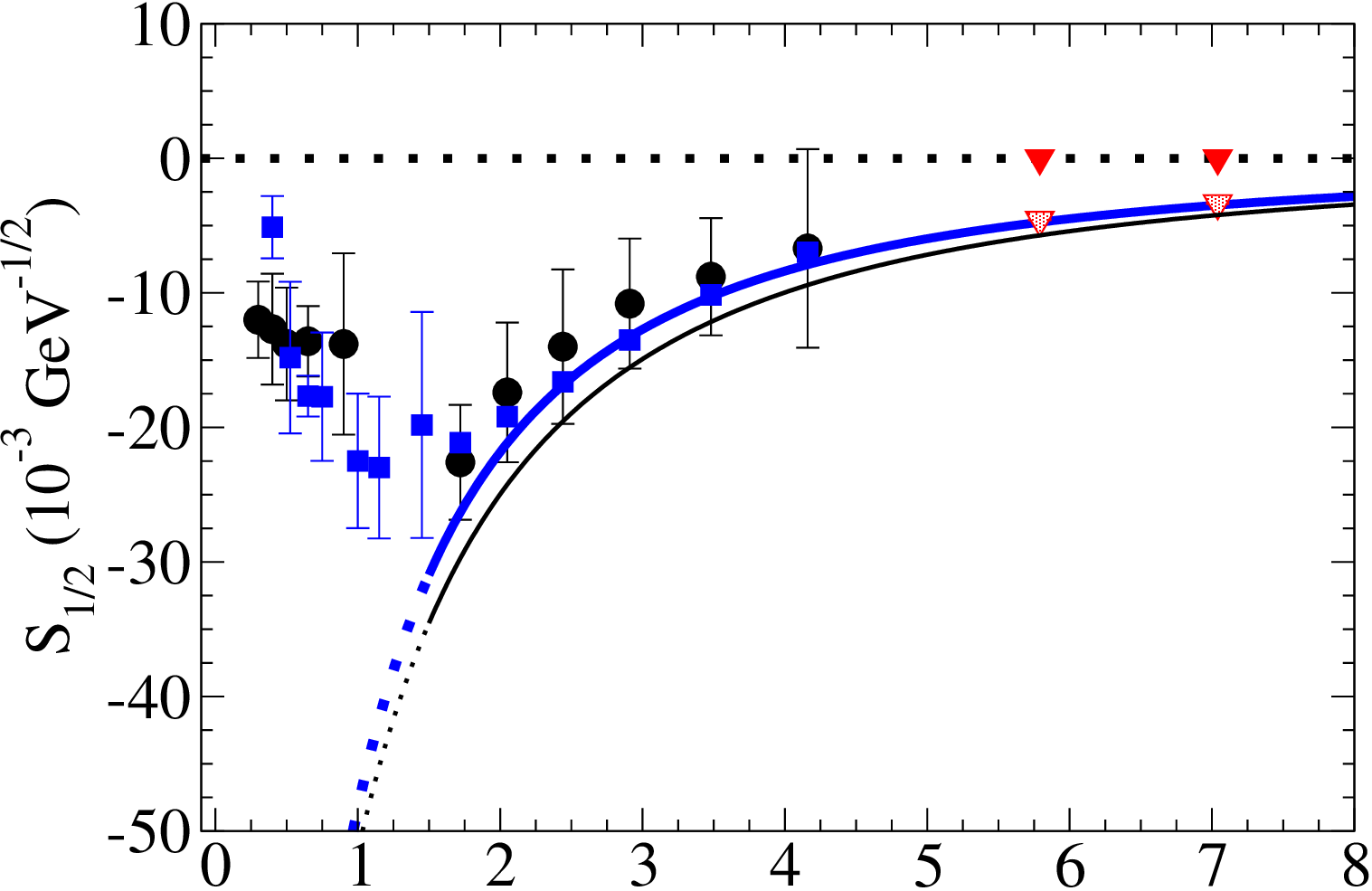}
}}
\centerline{
\vspace{.45cm} }
\centerline{
\mbox{
\includegraphics[width=2.8in]{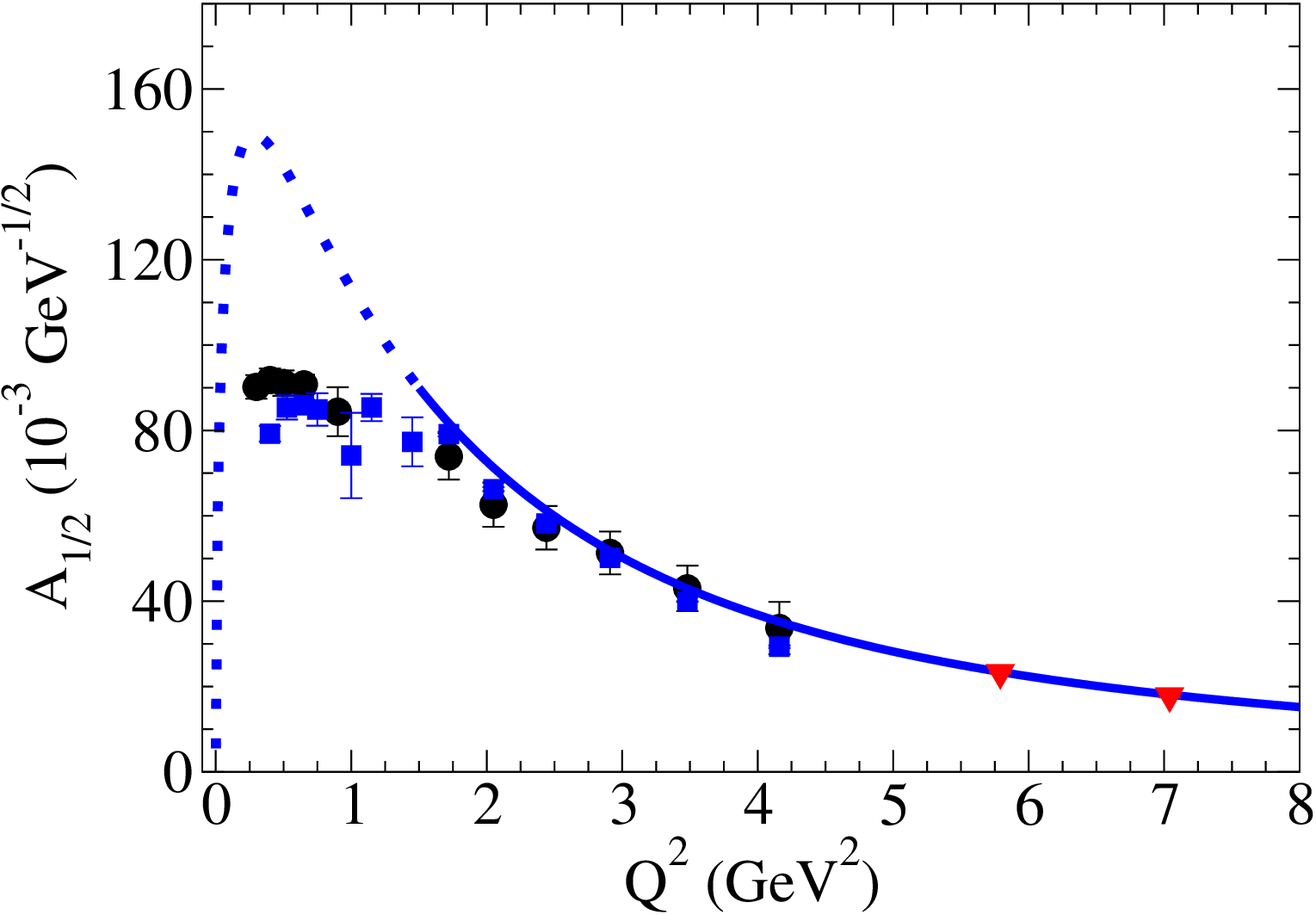}
}}
\caption{\footnotesize{
Results for the 
$\gamma^\ast N \to N(1535)$ helicity amplitudes given 
by the model B (thick solid line).
Model B is  valid for large $Q^2$ (see description in the main text). 
The thin solid line is the result of Eq.~(\ref{eqS12t}).
The dots represent the functions for $Q^2 < 1.5$ GeV$^2$.
Data from PDG (empty squares) \cite{PDG}, CLAS \cite{CLAS1} 
(full circles), MAID \cite{MAID1,MAID2} 
(full squares),
Jlab/Hall C \cite{Dalton09} (triangles).  
}}
\label{figS11amps2}
\end{figure}

The results for the amplitudes  
are presented in Fig.~\ref{figS11amps1} for the model A,
and in Fig.~\ref{figS11amps2} for the model B.
In the figures, we include the data from Ref.~\cite{Dalton09}
for $Q^2> 5$ GeV$^2$.
In the case of $S_{1/2}$, we include also 
the estimate from Ref.~\cite{S11b}, 
as discussed earlier (empty triangles).
In the graphs for $S_{1/2}$ the thick lines 
represent the original result for the amplitude,
given by Eq.~(\ref{eqS12X}) 
and the thin line the redefinition (\ref{eqS12t}).
In the last case, $S_{1/2}(0)$ is finite, 
although it is not shown in the graph.

In Fig.~\ref{figS11amps1}, we compare 
the valence quark model (model A) 
with the physical data (PDG, CLAS, MAID and JLab/Hall C).
In the graph for $S_{1/2}$ one can notice 
the significant disagreement between the model and the data.
The model strongly underestimates the data,
particularly for small values of $Q^2$.
Since the model A is based on valence quark contributions,
the deviation from the physical data 
may be an indication of the large meson cloud
effect expected for the amplitude $S_{1/2}$.
As for the amplitude $A_{1/2}$ it may be a surprise 
to see that the model is so close to the physical data,
since the model fails to describe the  $F_2^\ast$ data 
(see Fig.~\ref{figS11FF}).
This result is a consequence of 
the difference between model and data 
for the form factors $F_1^\ast$ and $F_2^\ast$, for $Q^2 < 2$ GeV$^2$,
combined with the difference of sign 
between $F_1^\ast$ and $F_2^\ast$, 
in the calculation of $A_{1/2}$ given by Eq.~(\ref{eqA12X}).
The closeness between the model A and the $A_{1/2}$ data 
for small $Q^2$ may be interpreted as a coincidence 
due to the results observed for the form factors 
(the model cannot describe simultaneously 
the form factors $F_1^\ast$, $F_2^\ast$ 
in the region $Q^2 < 2$ GeV$^2$).
As for large $Q^2$ the 
closeness between  model and data 
is expectable  due to the predicted falloff 
of the meson cloud contributions,
and also due to the smaller impact of 
$F_2^\ast$ in $A_{1/2}$ [see Eq.~(\ref{eqA12X})].

The model A may be useful 
in the future to compare with lattice QCD simulations 
with large pion  masses (small meson cloud effects)
and other estimates of the baryon core effects 
as the ones performed by the 
EBAC/Argonne-Osaka model~\cite{EBACreview,EBAC,Kamano16}
In future works one may also use the difference between 
our estimate of the bare core 
and a parametrization of the data to 
extract the contributions of the meson cloud.

In Fig.~\ref{figS11amps2}, we compare 
the high $Q^2$ model (model B)
directly with the data.
Since the result $F_2^\ast=0$ 
is observed only for $Q^2> 1.5$ GeV$^2$,
we represent the lines differently 
below (dotted line) and above (solid line) that point.
For $Q^2 > 1.5$ GeV$^2$ it is clear 
the agreement between the model and the physical data
(CLAS, MAID and JLab/Hall C) for both amplitudes.
In the graph for $S_{1/2}$ 
the results from Eq.~(\ref{eqS12X}), 
which diverge at $Q^2=0$ (thick line),  are closer 
to the data than the result from Eq.~(\ref{eqS12t}) (thin line), 
and are finite at $Q^2=0$. 
Both estimates are very close to the data 
in the region of interest ($Q^2> 1.5$ GeV$^2$).

The closeness between the model B and the data for $Q^2> 1.5$ GeV$^2$
is very interesting 
and calls for additional experimental studies,  
in order to test 
the hypothesis $F_2^\ast =0$ in more detail.
Also noticeable is the agreement 
between the model and the  estimate 
of the $S_{1/2}$ amplitude from Ref.~\cite{S11b} 
(empty triangles)
using the data from JLab/Hall C \cite{Dalton09}.
To clarify this point,  
new data or a reanalysis of old 
data using the Rosenbluth separation method 
(that allows the separation of different components 
of the measured cross section) may be very helpful.



\subsection{Discussion}

In the previous sections, we improved the results of the 
covariant spectator quark model from Refs.~\cite{D13,S11}
using the semirelativistic approximation.
The orthogonality between states is ensured
and the analytic results are consistent with the low $Q^2$ data.
In particular, we obtain nonzero results 
for $A_{1/2}$, $G_M$, $G_E$ at $Q^2=0$ in the 
$\gamma^\ast N \to N(1520)$ transition, and 
preserve the result $F_1^\ast(0) =0$ 
for the  $\gamma^\ast N \to N(1535)$ transition.

Compared to the  models from Refs.~\cite{D13,S11},
where the estimate of the valence quark contributions
for the form factors were good only for large $Q^2$,
we present more reliable estimates for the low $Q^2$ region, 
although derived under the assumption that $M_R \simeq M_N$.
An accurate estimate of the valence quark contributions 
in the low $Q^2$ regime is very important, 
since it can be used to estimate the meson cloud contributions
based on a parametrization of the form factor data.
A parametrization of the 
valence quark contributions can also be very 
useful to compare with lattice simulations with large pion masses 
(suppression of meson cloud effects)
and other estimates of the bare core contributions.

Our results for the $\gamma^\ast N \to N(1520)$ transition 
are in good agreement with the intermediate 
and large $Q^2$ data ($Q^2 > 1$ GeV$^2$).
The exceptions are the amplitude $A_{3/2}$, 
for which the spectator quark model 
predicts zero contributions, and the form factor $G_E$.

The experimental results (with meson cloud) 
and the estimates of bare core effects,
such as the one based on the semirelativistic approximation 
(without meson cloud) for $G_E$, can be understood in the case where 
$A_{3/2}$ is mainly a consequence 
of the  meson cloud effects 
and $A_{1/2}$ is dominated by valence quark effects 
(small meson cloud contributions).
The relations between the meson cloud (index mc) contributions 
from the helicity amplitudes and form factors
can be represented as~\cite{Aznauryan12a,D13}
\ba
& &
A_{1/2}^{\rm mc}= 
\frac{1}{4F} (3 G_M^{\rm mc} - G_E^{\rm mc}) ,  \\
& &
A_{3/2}^{\rm mc}= 
 - \frac{\sqrt{3}}{4F} (G_M^{\rm mc} + G_E^{\rm mc}), 
\ea
where $F={\cal R}/(2 {\cal A}_R)$ is a function of $Q^2$~\cite{D13}.
When $|A_{3/2}^{\rm mc}| \gg |A_{1/2}^{\rm mc}|$,
we can conclude that
\ba
G_E^{\rm mc} = - \frac{F}{\sqrt{3}} A_{3/2}^{\rm mc},
\hspace{.8cm}
G_M^{\rm mc} = \sfrac{1}{3}G_E^{\rm mc}.
\ea
Thus, in the case where the meson cloud contributions are 
large for $A_{3/2}$ and small for $A_{1/2}$,
$G_E$ has larger meson cloud contributions, 
proportional to $A_{3/2}^{\rm mc}$, 
and $G_M$  has smaller meson cloud contributions
(about one third of the contribution for $G_E$).
Those results are compatible 
with the results of Figs.~\ref{figD13amps} 
and \ref{figD13FF} for $A_{1/2}$ and $G_M$.
It is worth mentioning that the dominance of the meson cloud effects 
in the amplitude $A_{3/2}$ 
was already observed in some EBAC calculations~\cite{EBAC}.
Indications of the the large meson cloud contributions 
for $A_{3/2}$ came also from quark models 
where, as mentioned, 
the valence quarks contribute only with  
a small fraction of the experimental 
values~\cite{Warns90,Aiello98,Merten02,Santopinto12,Ronniger13}.

Our results for the $\gamma^\ast N \to N(1535)$
for  $F_1^\ast$ are compatible with the experimental data
in the region $Q^2 > 1.5$ GeV$^2$, 
and differ in sign for $F_2^\ast$.
We can interpret those results as 
a manifestation of the absence of meson cloud effects,
particularly for $F_2^\ast$.
For the  $\gamma^\ast N \to N(1535)$  transition
the semirelativistic approximation with $M_R=M_N$
is unappropriated for the calculation 
of the helicity amplitudes.
One can use, however, two simple approximations: 
one based on the valence quark contributions (model A),
and another that is valid for large $Q^2$, 
and compares well with the data 
(model B, with $F_2^\ast=0$).

Overall, we conclude that we have a good description of the 
valence quark content of the $N(1520)$ and $N(1525)$ systems, 
since we describe very well the large $Q^2$ data.

Estimates of the valence quark contributions 
to the form factors can by performed using 
dynamical coupled-channel reaction models 
like the DMT~\cite{MAID1,Kamalov01}, 
and the EBAC/Argonne-Osaka model~\cite{EBAC,EBACreview,Kamano16}.
Those models take into account the 
meson and photon coupling with the baryon cores 
and can be used to estimate the effect of 
the bare core, when the meson cloud effects are removed,
or the effect of the meson cloud, 
when the bare core effect is subtracted~\cite{Burkert04,EBAC,EBACreview}.

The EBAC model has been in the past applied to the 
analysis of the CLAS data from Refs.~\cite{CLAS-1,CLAS-2,CLAS-3,CLAS-4},
including  $\gamma^\ast p \to \pi^+ n$
and $\gamma^\ast p \to \pi^0 p$ data,
and used to calculate the bare contributions 
for the $\gamma^\ast N \to N(1520)$ and $\gamma^\ast N \to N(1535)$
transition form factors~\cite{EBAC}.
At the time the analysis was restricted to $Q^2=0.4$ GeV$^2$. 
It was concluded that the analysis 
of the $\gamma^\ast p \to \pi^+ n$ data~\cite{CLAS-1} 
can differ significantly from the 
combined analysis of the  $\gamma^\ast p \to \pi^+ n$ and 
$\gamma^\ast p \to \pi^0 p$ data~\cite{CLAS-1,CLAS-2,CLAS-3,CLAS-4}.

We then expect that in the near future combined 
analysis of the  $\gamma^\ast p \to \pi^+ n$ and 
$\gamma^\ast p \to \pi^0 p$ data become available 
for a wide range of $Q^2$, in order to test 
our estimates of the valence quark contributions 
for the $\gamma^\ast N \to N(1520)$ and 
$\gamma^\ast N \to N(1535)$ transition form factors.


\section{Outlook and conclusions}  
\label{secConclusions}

In this work we present a new method to calculate 
the $\gamma^\ast N \to R$ transition form factors, 
where $R$ is a negative parity resonance,
within the covariant spectator quark model.
The method is named as the semirelativistic approximation,
since the nucleon and resonance wave functions
are defined both for the mass $M=\sfrac{1}{2}(M_N + M_R)$.

In the semirelativistic approximation
the properties of the nonrelativistic limit of the states, 
in particular the orthogonality 
between those wave functions and the nucleon wave function, 
are preserved, but the formalism is still covariant.
The wave functions of the $R$ states are  
defined using the same parametrization 
for the radial wave functions as for the nucleon.

We use analytic results from previous works 
and apply the semirelativistic approximation 
to the cases $R=N(1520)$, $N(1535)$.
Within the approximation we calculate 
the valence quark contributions for the 
transition form factors and helicity amplitudes.
Since the wave functions of those states are 
defined in terms of the parametrization for the nucleon, 
the method provides predictions 
for the transition form factors and helicity amplitudes.

In general, our estimates based exclusively 
on the valence quark degrees of freedom, 
are in excellent agreement with the results for the 
form factors in the region $Q^2 > 2$ GeV$^2$, 
where we expect very small contributions from the meson cloud.
We then conclude that we have a good description 
of valence quark content of 
the nucleon, $N(1520)$ and $N(1535)$ systems.
Our valence quark parametrizations can be compared 
in a near future with the bare contributions 
estimated from the combined analysis 
of the $\gamma^\ast p \to \pi^0 p$ and 
$\gamma^\ast p \to \pi^+ n$ data,
and with lattice QCD simulations.

The semirelativistic approximation is more appropriated 
for the form factors than for the helicity amplitudes.
The calculation of the helicity amplitudes 
must be done with some care, 
since it depends critically on the mass difference,
particularly in the  $\gamma^\ast N \to N(1535)$ case.
For the $\gamma^\ast N \to N(1520)$ transition 
we obtained a very good description 
of the  helicity amplitudes for $Q^2 > 1$ GeV$^2$.
The amplitude $A_{3/2}$ is the exception,
since this amplitude is expected to be dominated by meson cloud effects.
As for the $\gamma^\ast N \to N(1535)$ transition, 
we present parametrizations of the amplitudes valid for large $Q^2$.

From our study, we can conclude that the $N(1520)$ and $N(1535)$ 
resonances are very interesting physical systems.
The transition form factors associated 
with the $N(1520)$ and $N(1535)$ resonances are in general 
dominated by the the valence quark effects with a few exceptions.
The electric form factor $G_E$ in the $\gamma^\ast N \to N(1520)$ transition is  
strongly dominated by meson cloud contributions.
There is also evidence that the form factors 
$F_1^\ast$ and $F_2^\ast$ in the $\gamma^\ast N \to N(1535)$ transition
have important meson cloud contributions.
The effect on $F_2^\ast$ was discussed already in the literature.

To summarize, we present parametrizations 
for the $\gamma^\ast N \to N(1520)$ and $\gamma^\ast N \to N(1535)$
transition form factors and respective helicity amplitudes, 
that are consistent with the available data 
in the regime of $Q^2 = 2$--7 GeV$^2$.
Our predictions may be tested in the future JLab 12-GeV upgrade 
up to 12 GeV$^2$~\cite{Jlab12GeV}.
Of particular interest is the test of our high $Q^2$ model 
 for the $\gamma^\ast N \to N(1535)$ transition 
(with $F_2^\ast=0$), which predicts 
the relation between amplitudes:
$S_{1/2} = - \frac{\sqrt{1+ \tau}}{\sqrt{2}} \frac{M_R^2 -M_N^2}{2 M_R Q} A_{1/2}$ 
\cite{S11b} and was till the moment tested 
only up to $Q^2=4.2$ GeV$^2$
by the CLAS and MAID data \cite{CLAS1,MAID1,MAID2}.

\vspace{.2cm}

\begin{acknowledgments}
The author thanks Hiroyuki Kamano for helpful discussions.
This work was supported by the Brazilian Ministry of Science,
Technology and Innovation (MCTI-Brazil).
\end{acknowledgments}


\begin{thebibliography}{00}
\bibitem{Note} 
   In comparison to Refs.~\cite{S11,S11b} 
the form factors are corrected by a sign.
\end{thebibliography}
\end{document}